\documentclass[vecphys]{svmult}

\usepackage{makeidx}         
\usepackage{graphicx}        
\usepackage{multicol}        
\usepackage{cite}            
\usepackage[bottom]{footmisc}
\usepackage{amssymb,amsmath}
\usepackage{graphicx}

\newcommand{\Felec}{F_\mathrm{elec}}
\newcommand{\Fexc}{F_\mathrm{exc}}
\newcommand{\Fts}{F_\mathrm{ts}}
\newcommand{\Vbias}{V_\mathrm{bias}}
\newcommand{\Vcpd}{V_\mathrm{cpd}}

\newcommand{\dcdz}{\frac{\partial C}{\partial z}}

\newcommand{\Vdc}{V_\mathrm{dc}}
\newcommand{\Vac}{V_\mathrm{ac}}

\newcommand{\zO}{z_\mathrm{0}}

\newcommand{\wO}{\omega _\mathrm{0}}
\newcommand{\df}{\Delta f}

\newcommand{\Fdc}{F_\mathrm{dc}}
\newcommand{\Fw}{F_\mathrm{\omega}}
\newcommand{\Fww}{F_\mathrm{2\omega}}

\newcommand{\Fin}{F_\mathrm{in}}
\newcommand{\Fout}{F_\mathrm{quad}}
\newcommand{\wm}{\omega_\mathrm{m}}
\newcommand{\wel}{\omega_\mathrm{el}}

\makeindex             

\begin{document}

\title{Dissipation modulated Kelvin probe force microscopy method}
\author{Yoichi Miyahara \and Peter Grutter}
\institute{Department of Physics, McGill University
\texttt{yoichi.miyahara@mcgill.ca}}

\maketitle

\begin{abstract}
We review a new experimental implementation of Kelvin probe force
microscopy (KPFM) in which the dissipation signal of frequency modulation 
atomic force microscopy (FM-AFM) is used for dc bias voltage feedback (D-KPFM).
The dissipation arises from an oscillating electrostatic force 
that is coherent with the tip oscillation,
which is caused by applying the ac voltage between the tip and sample.
The magnitude of the externally induced dissipation is found to be proportional
to the effective dc bias voltage, 
which is the difference between the applied dc voltage 
and the contact potential difference.
Two different implementations of D-KPFM are presented. 
In the first implementation, the frequency of the applied ac voltage, 
$f_\mathrm{el}$, is chosen to be the same as the tip oscillation
($f_\mathrm{el} = f_\mathrm{m}$: $1\omega$D-KPFM).
In the second one, the ac voltage frequency, $f_\mathrm{el}$, is chosen to be 
twice the tip oscillation frequency 
($f_\mathrm{el}= 2 f_\mathrm{m}$: $2\omega$D-KPFM).
In $1\omega$D-KPFM, the dissipation is proportional to the electrostatic force,
which enables the use of a small ac voltage amplitude 
even down to $\approx 10$\,mV.
In $2\omega$D-KPFM, the dissipation is proportional 
to the electrostatic force gradient,
which results in the same potential contrast as that obtained 
by FM-KPFM.
D-KPFM features a simple implementation with no lock-in amplifier
and faster scanning as it requires no low frequency modulation.
The use of a small ac voltage amplitude in $1\omega$D-KPFM is of great importance 
in characterizing of technically relevant materials
in which their electrical properties can be disturbed 
by the applied electric field. 
$2\omega$D-KPFM is useful when more accurate potential measurement is
required. 
The operations in $1\omega$ and $2\omega$D-KPFM can be switched easily 
to take advantage of both features at the same location on a sample.
\end{abstract}

\noindent
\section{Introduction}
\label{sec:intro}
Kelvin probe force microscopy (KPFM), a variant of atomic force microscopy
(AFM) has become one of the indispensable tools 
used to investigate electronic properties of nanoscale material as well as
nanoscale devices.
In KPFM, a contact potential difference (CPD) between the AFM tip and sample surface 
is measured by detecting a capacitive electrostatic force, $\Felec$, 
that is a function of the CPD, $\Vcpd$, and applied bias voltage, 
$\Vbias$ such as $\Felec \propto (\Vbias - \Vcpd)$.
In order to separate the electrostatic force component from other force 
components such as van der Waals force, chemical bonding force 
and magnetic force, 
the capacitive electrostatic force is modulated by applying an ac voltage.
and the resulting component of the measured observable,
which are typically the resonant frequency shift or the change in amplitude 
of an oscillating AFM cantilever, 
is detected by lock-in detection~\cite{Nonnenmacher91}.

KPFM has been implemented in a variety of ways that can be classified mostly
into two distinct categories, amplitude modulation (AM-) KPFM
~\cite{Nonnenmacher91, Kikukawa96, Sommerhalter1999, Zerweck05} 
and frequency modulation (FM-) KPFM~\cite{Kitamura98, Zerweck05}.

In AM-KPFM, one of the cantilever resonance modes is excited by applying 
an ac voltage 
and the resulting oscillation amplitude is detected 
as a measure of the capacitive electrostatic force,
which is used for controlling the dc bias voltage to nullify the oscillation amplitude.
This implementations can take advantage of enhanced electrostatic
force detection sensitivity 
by tuning the modulation frequency to one of the resonance frequencies of 
the AFM cantilever, leading to an enhanced detection of the electrostatic force 
by its quality ($Q$) factor that can reach over 10000 in vacuum environment~\cite{Schumacher2015}.
In single-pass implementation in which the topography and CPD images are taken
simultaneously, the second flexural mode is usually chosen for detecting
the electrostatic force 
while the first flexural mode is used for detecting short-range interaction 
which is used for the topography imaging.

In FM-KPFM, a low frequency (typically several hundred Hz) 
ac voltage is superposed with the dc bias voltage,
resulting in the modulation in the resonance frequency shift.
The amplitude of the modulated resonance frequency shift is detected 
by a lock-in amplifier and then used for the bias voltage feedback.
Although this method requires a much higher ac voltage amplitude than AM-KPFM, 
it offers higher spatial resolution because the resonance frequency shift is 
determined by the electrostatic force gradient 
with respect to the tip-sample distance
rather than the force itself~\cite{Sommerhalter1999, Zerweck05, Burke09a}.

In this chapter, we report two alternative KPFM implementations
($1\omega$D-KPFM and $2\omega$D-KPFM) in which the dissipation signal 
of a frequency modulation atomic force microscopy (FM-AFM)
is used for detecting the capacitive electrostatic force~\cite{Miyahara2015, Miyahara2017b}.
The dissipation is induced by applying a coherent sinusoidal ac voltage 
which is $90^\circ$ out of phase with respect to the tip oscillation.
The externally induced dissipation signal can be used for 
the dc bias voltage feedback
as it is proportional to the effective dc potential difference, 
$\Vdc \equiv (\Vbias - \Vcpd)$.

In $1\omega$D-KPFM, the angular frequency of the applied ac voltage, 
$\wel$, is chosen to be the same as the tip oscillation
($\wel = \wm$).
In $2\omega$D-KPFM, $\wel$ is chosen to be twice $\wm$
($\wel= 2 \wm$).
We will show that, in $1\omega$D-KPFM, the induced dissipation 
is proportional to the electrostatic force,
which enables the use of a small ac voltage amplitude down to $\approx 10$\,mV
whereas, in $2\omega$D-KPFM, the dissipation is proportional 
to the electrostatic force gradient,
which results in the same potential contrast as that obtained 
by FM-KPFM.

\begin{figure}[t]
\centering
\includegraphics[width=.7\textwidth]{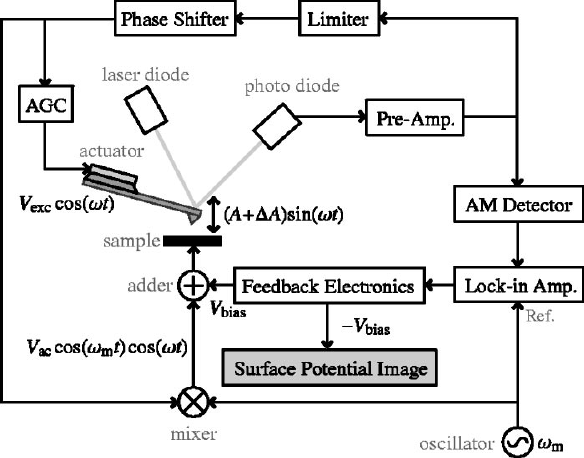}
\caption[]{Block diagram of dissipation modulated Kelvin probe force 
microscopy technique proposed by Fukuma \textit{et al.}~\cite{Fukuma2004}.
The induced dissipation is detected by measuring the change in the 
amplitude of tip oscillation.
Reprinted from~\cite{Fukuma2004}, with the permission of AIP Publishing}
\label{fig:Fukuma_diagram}       
\end{figure}

\begin{figure}[]
\centering
\includegraphics[width=.5\textwidth]{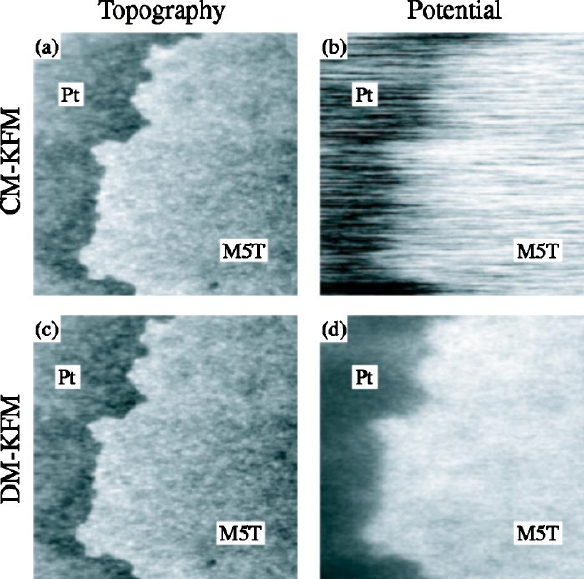}
\caption[]{Topography and CPD images of dimethylquinquethiophene monolayer formed 
on a Pt surface. (a) topography and (b) CPD images taken by FM-KPFM.
(c) topography and (d) CPD images taken by DM-KPFM. From \cite{Fukuma2004}.
Reprinted from~\cite{Fukuma2004}, with the permission of AIP Publishing}
\label{fig:Fukuma_image}       
\end{figure}

The idea of using induced dissipation for KPFM was first reported 
by Fukuma \textit{et al.}~\cite{Fukuma2004} which they named DM-KPFM.
In their implementation, the dissipation is measured through the change 
in tip oscillation amplitude rather than the dissipation signal (Fig.~\ref{fig:Fukuma_diagram}).
Despite the demonstrated higher sensitivity of DM-KPFM (Fig.~\ref{fig:Fukuma_image}),
it has not been widely adopted, probably because of its rather complex
implementation and limited detection bandwidth due to the amplitude 
detection~\cite{Fukuma2004}.
The use of the ac voltage with twice the frequency of the tip oscillation 
was proposed by Nomura \textit{et al.}~\cite{Nomura2007} for DM-KPFM
to be sensitive to electrostatic force gradient rather than electrostatic force itself.

Our implementation of D-KPFM features a simple implementation 
with no lock-in amplifier
enabling faster scanning as it requires no low frequency modulation.
The use of a small ac voltage amplitude in $1\omega$D-KPFM is of great importance 
in characterizing technically relevant materials
in which their electrical properties can be disturbed 
by the applied electric field 
(\textit{e.g.} resulting in band-bending.effects at interfaces).
$2\omega$D-KPFM is useful when more accurate potential measurement is
required. 
The operations in $1\omega$ and $2\omega$D-KPFM can be switched easily 
to take advantage of both features at the same location on a sample.

\section{Theory}
In order to understand the operation principle of D-KPFM technique,
we will first review the theory of FM-AFM and the effect of a periodic 
applied force on the in-phase and out-of-phase signal
(commonly known as frequency shift and dissipation) 
of the FM-AFM system.
We will then discuss the detailed analysis of the electrostatic force 
in the presence of the applied coherent ac voltage
and how its effect appears in the resonant frequency shift and dissipation.

\subsection{Review of theory of frequency modulation 
atomic force microscopy }
\label{sec:FM-AFM theory}
\paragraph{Frequency shift and Dissipation 
in frequency modulation atomic force microscopy}
We consider the following equation of motion of the AFM cantilever to model FM-AFM.
\begin{equation}
  \label{eq:eom}
  m\ddot{z}(t) + m \frac{\wO}{Q} \dot{z}(t) + k \times (z(t)-\zO)= \Fts(t) + \Fexc(t)
\end{equation}
where $m$, $\wO$, $Q$, $k$ are the effective mass and angular resonance frequency,
mechanical quality factor and effective spring constant of the AFM cantilever.
$\zO$ is the mean distance of the tip measured from the sample surface.
$\Fts(t)$ and $\Fexc(t)$ are the force acting on the tip caused by tip-sample
interaction
and an external drive force that 
is used to excite the cantilever oscillation, respectively.
In FM-AFM, the AFM cantilever is used as a mechanical resonator 
with a high $Q$ (typically $> 1000$) 
which acts as a frequency determining component
of a self-driven oscillator~\cite{Albrecht1991}. 
Such a self-driven oscillator is realized by a positive feedback circuit 
equipped with an amplitude controller 
that keeps the oscillation amplitude constant~\cite{Albrecht1991, Durig1997}.
In this case, the external driving force, $\Fexc(t)$, is generated
by a time-delayed (phase-shifted) cantilever deflection 
such as $\Fexc(t)=g k z(t-t_0)$
which is commonly transduced 
by piezoacoustic or photothermal excitation scheme~\cite{Labuda2012a}
through a phase shifter electronics.
Here $g$ represents the gain of the positive feedback circuit and is called 
\textit{dissipation signal},
and $t_0$ is the time delay set by the phase shifter~\footnote{In general, 
$g$ takes a complex value 
if the transfer function of the excitation system is to be taken into account.
We neglect the effect of the transfer function here.
See Ref.~\cite{Labuda2011} for more detail.}.
For the typical cantilever with a high $Q$ factor and high spring constant
used for FM-AFM, 
we can assume a harmonic oscillation of the cantilever 
such as $z(t)=z_0 + A \cos(\wm t)$.
When $t_0$ is set to be $\frac{1}{4}T_0, \frac{3}{4}T_0, \frac{5}{4}T_0, \cdots$
($T_0 \equiv 2\pi/\wO$), 
the oscillation frequency, $f_\mathrm{m}=\wm/2\pi$, tracks 
its mechanical resonance frequency such that $f_\mathrm{m} = f_0 (\equiv \wO/2\pi)$.
In this condition, the frequency shift, $\Delta f$, and dissipation signal, $g$,
are expressed as follows~\cite{Holscher2001,Kantorovich2004, Sader2005}:
\begin{eqnarray}
       \df & \approx & -\frac{f_0^2}{kA}\int^{T_0}_0 \Fts(t) \cos(\wm t)dt   \label{eq:df}\\
        g & \approx & \frac{1}{Q}+\frac{2 f_0^2}{kA} \int^{T_0}_0 \Fts(t) \sin(\wm t)dt \label{eq:g}.
\end{eqnarray}
It is important to notice that in general, 
the force acting on the tip, $\Fts(t)$, can have
an explicit time dependence 
in addition to the time dependence due to the time-varying tip position 
which is expressed as $\Fts(z(t))$~\cite{Kantorovich2004}.
The explicit time dependence can originate from various tip-induced processes 
such as dynamic structural relaxation of either tip 
or/and sample~\cite{Kantorovich2004} and single-electron tunneling~\cite{Miyahara2017}.
In the case of electrostatic force,
$\Fts(t)$ is determined by the applied voltage as well as the tip position.
It is thus essential to take into account both dependencies explicitly.
Before going into further detail of the electrostatic force,
we will look at the effect of time-varying periodic force 
on frequency shift and damping of the cantilever. 

\paragraph{Effect of coherent periodic force on frequency shift and dissipation}
\label{sec:Effect_of_coherent_force}

\begin{figure}[t]
  \centering
   \includegraphics[width=70mm]{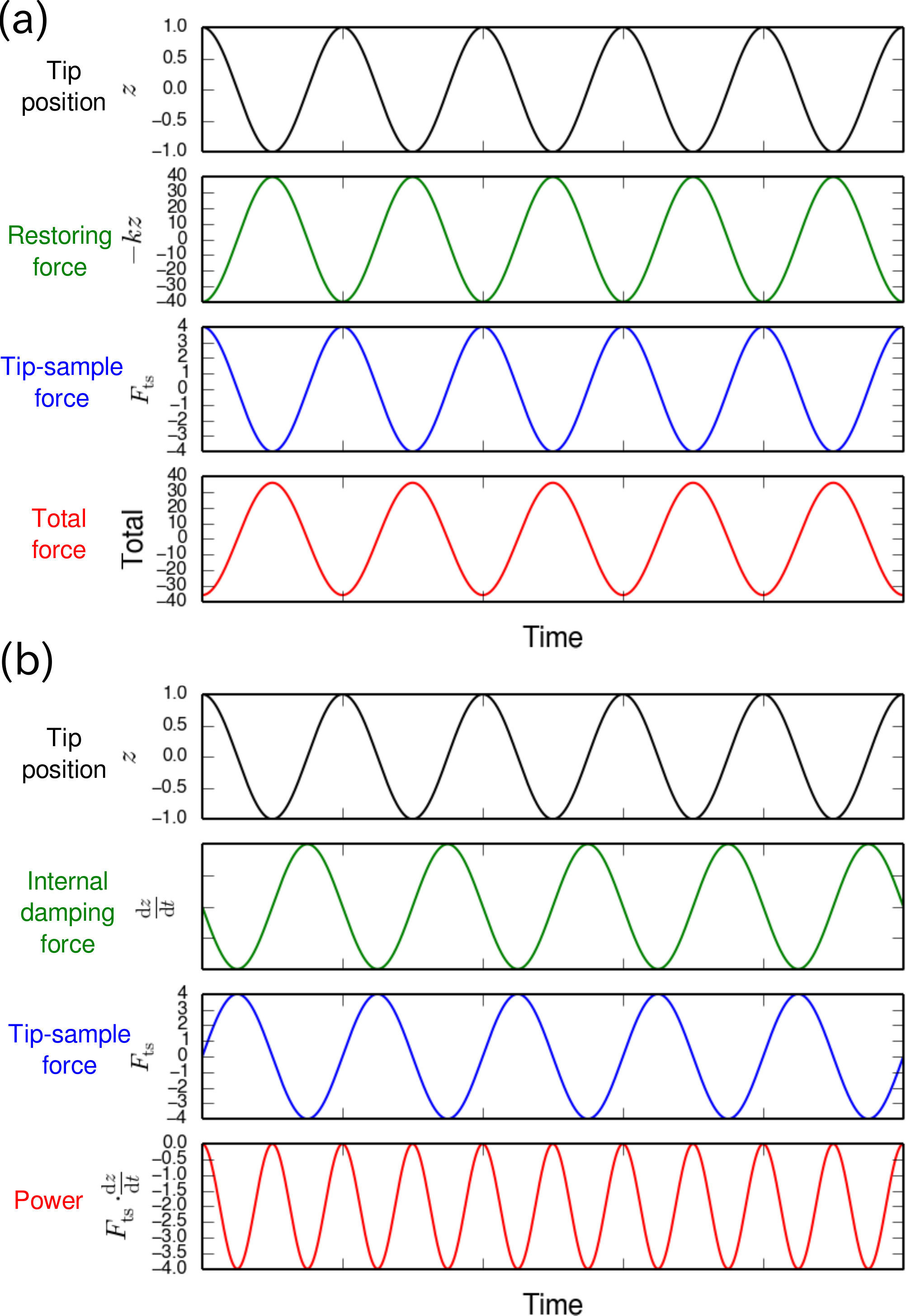}
\caption[]{(a) Schematic representation of tip position, $z$, 
restoring force, $-kz$, 
the in-phase fundamental Fourier component, $\Fin \cos(\wm t)$,
and the total force.
(b) Schematic representation of tip position, $z$, 
internal damping force, $m\gamma_0 \dot{z}$, 
the quadrature fundamental Fourier component, $\Fout \sin(\wm t)$,
and the instantaneous power delivered from the tip-sample interaction
}
\label{fig:effect_of_periodic_f}
\end{figure}

In general, the periodically oscillating force, $\Fts(t)$, 
 can be represented by a Fourier series:
\begin{equation}
  \label{eq:Fouier_series}
  \Fts(t) = F_0 + \sum_{n=1} ^{\infty} \{\Fin^{(n)} \cos (n\wm t) + \Fout^{(n)} \sin (n\wm t)\} 
\end{equation}
The cosine terms represent the force component which is in phase (even) with 
the tip oscillation, which is conservative
and the sine terms represent $90^\circ$ out-of-phase (quadrature, odd) component 
which is dissipative~\cite{Sader2005}.
Substituting Eq.~\ref{eq:Fouier_series} into Eq.~\ref{eq:df} and \ref{eq:g} 
yields the following results \footnote{We use the identities, 
$\displaystyle \int_{0} ^{T_0} \cos (n\wm t) \cos(\wm t)=0 $,
and $\displaystyle \int_{0} ^{T_0} \sin (n\wm t) \sin(\wm t)=0$ 
for $n \neq 1$.
}:
\begin{eqnarray}
  \Delta f & = & -\frac{1}{2} \frac{f_0}{kA} \Fin^{(1)} \label{eq:df_Fin}\\
  g & = & g_0 \left(1+\frac{Q}{kA}\Fout^{(1)}\right) \label{eq:g_Fout}
\end{eqnarray}
where $g_0 \equiv 1/Q$ is the dissipation signal in the absence of $\Fts(t)$,
which is determined by the intrinsic damping of the cantilever.
Eq.~\ref{eq:df_Fin} and \ref{eq:g_Fout} indicate that $\df$ and $g$ are proportional to 
the Fourier in-phase (even), $\Fin (\equiv \Fin^{(1)})$, and quadrature (odd), 
$\Fout (\equiv \Fout^{(1)})$, coefficients of the fundamental harmonic component 
(in this case $\wm$) of $\Fts(t)$, respectively. 
We can understand this results intuitively with the schematics shown 
in Fig.~\ref{fig:effect_of_periodic_f}.
Fig.~\ref{fig:effect_of_periodic_f}(a) shows the tip position, $z$, 
restoring force, $-kz$,  
the in-phase fundamental Fourier component, $\Fin \cos(\wm t)$,
and the total force acting on the cantilever.
Similarly, Fig.~\ref{fig:effect_of_periodic_f}(b) shows the tip position, $z$, 
intrinsic damping force, $m\gamma_0 \dot{z}$, 
the quadrature fundamental Fourier component, $\Fout \sin(\wm t)$,
and the instantaneous power delivered from the tip-sample interaction.
As can be seen from these figures, 
while the in-phase component just influences the restoring force, resulting in 
the resonance frequency shift,
the quadrature component can change the damping force,
resulting in a signal in the dissipation channel.
This can also be understood by non-zero average power delivered by the quadrature tip-sample
interaction as shown in the power-time plot (Fig.~\ref{fig:effect_of_periodic_f}(b)).

\subsection{Analysis of electrostatic force with ac bias voltage}
The electrostatic force between two conductors connected 
to an ac and dc voltage source is described as follows:
\begin{eqnarray}
      \Felec (t) & = & \frac{1}{2}\dcdz \{\Vbias +\Vac \cos (\wel t + \phi) - \Vcpd\}^2 \\
      & = & \Fdc + \Fw + \Fww \nonumber \\
       F_0 & = & \frac{1}{2} \dcdz \left[(\Vbias - \Vcpd)^2 + \frac{\Vac ^2}{2}\right] \label{Fdc} 
       = \alpha \left(\Vdc ^2 + \frac{\Vac^2}{2} \right)\\ 
       \Fw &= &  \dcdz (\Vbias - \Vcpd)\Vac \cos (\wel t + \phi)\nonumber \\ 
       & = & 2\alpha \Vdc \Vac \cos (\wel t + \phi)  \label{Fw} \\ 
       \Fww & = & \frac{1}{4} \dcdz \Vac ^2 \cos \{2(\wel t + \phi)\}  \label{eq:F2w} 
       = \frac{1}{2}\alpha \Vac^2 \cos \{2(\wel t + \phi)\}
\end{eqnarray}
where $C$ is the tip-sample capacitance, $\Vbias$ and $\Vcpd$ are the applied dc voltage 
and the contact potential difference (CPD), 
and $\Vac$, $\wel$, $\phi$ are the amplitude, angular frequency, 
and phase of the applied ac bias voltage.
$\Vdc \equiv \Vbias -\Vcpd$ is the effective dc bias voltage.
It is important to notice that although the three force component terms, 
$F_0$, $\Fw$, $\Fww$, are grouped by the frequency of $\wel$ harmonic component
such as 0, $\omega$, $2\omega$,
they do not represent the frequency of harmonic components of the actual 
oscillating electrostatic force
in the presence of the mechanical tip oscillation
because $\displaystyle \alpha \equiv \frac{1}{2}\dcdz(z(t))$ is 
also a function of time through the time-dependent tip position 
$z(t)=z_0 + A\cos(\wm t)$.
It is the interaction of the oscillating electric field 
and mechanical tip oscillation
that results in an ``electro-mechanical heterodyning'' effect as we will see below.

By Taylor expanding $\alpha(z)$ around the mean tip position $z_0$,
\begin{eqnarray}
  \alpha (z) & = & \alpha(z_0) + \alpha ' (z-z_0) + \frac{1}{2}\alpha'' (z-z_0)^2 + .... \nonumber \\
  & = & \alpha_0 + \alpha ' A \cos(\wm t) + \frac{1}{2}\alpha'' A^2 \cos^2(\wm t) + ....
\end{eqnarray}

Taking the first order yields,
\begin{equation}
  \alpha (z) \approx \alpha_0 + \alpha ' (z-z_0) = \alpha_0 + \alpha ' A \cos(\wm t).
\end{equation}

Substituting $\alpha(z)$ into Eq.~\ref{Fdc} yields
\begin{eqnarray}
       F_0(t)  &= & \{\alpha_0 + \alpha' A \cos(\wm t)\} \left(\Vdc ^2 + \frac{\Vac^2}{2}\right) \nonumber \\
       &=& \alpha_0 \left(\Vdc ^2 + \frac{\Vac^2}{2}\right)  
       + \alpha' A\left(\Vdc ^2 + \frac{\Vac^2}{2}\right) \cos (\wm t).
      \label{eq:Fdc2} 
\end{eqnarray}
The first term in Eq.~\ref{eq:Fdc2} represents a time-invariant force 
which causes the static deflection of the cantilever 
whereas the second term represents an oscillating force with frequency of $\wm/2\pi$.
As this oscillating force is in phase with $z(t)$, 
it causes the shift in the resonance frequency.

Next, we look at $\Fw$ and $\Fww$ terms (Eq.~\ref{Fw} and \ref{eq:F2w})
in the same way:
\begin{equation}
  \begin{split}
    \label{eq:Fw2} 
       \Fw(t) &=  2 \{\alpha_0 + \alpha' A \cos(\wm t)\} \Vdc\Vac \cos (\wel t + \phi)\\
       & =  2\alpha_0 \Vdc\Vac \{\cos(\wel t)\cos\phi- \sin(\wel t)\sin\phi \}  \\
       & + \alpha' A\Vdc\Vac [\cos\{(\wel + \wm) t + \phi\} + \cos\{(\wel - \wm)t +\phi\}]  
\end{split}
\end{equation}
\begin{equation}
  \begin{split}
  \label{eq:F2w2}
       \Fww(t) &= \frac{1}{2} \{\alpha_0 +  \alpha' A \cos (\wm t) \} \Vac ^2 \cos \{2(\wel t + \phi)\}\\
       & =  \frac{1}{2}  \alpha_0 \Vac ^2 \cos \{2(\wel t + \phi)\} \\ 
       & + \frac{1}{4} \alpha' A \Vac ^2 [ \cos \{(2\wel - \wm)t + 2\phi\}
       +  \cos \{(2\wel + \wm) t + 2\phi\}].
     \end{split}
\end{equation}
Here, we notice that the electro-mechanical heterodyning produces other spectral components 
($\wel+\wm$, $\wel-\wm$, $2\wel-\wm$, $2\wel+\wm$) than 
$\wel$ and $2\wel$ which are usually considered in the most KPFM literature. 
In order to calculate the frequency shift and damping, 
these additional spectral components need to be included.
In the following, we consider two interesting cases (1) $\wel = \wm$, 
and (2) $\wel = 2\wm$ 
which corresponds to $1\omega$D-KPFM and $2\omega$D-KPFM, respectively.

\paragraph{Case I , $\wel = \wm $: $1\omega$D-KPFM}
\label{sec:1fDKPFM}
In the case of $\wel = \wm$,
gathering the time invariant terms found in $F_0$ (Eq.~\ref{eq:Fdc2}) 
and $\Fw$ (Eq.~\ref{eq:Fw2}), 
the actual dc term, $\Fdc (t)$ can be found to be: 
\begin{eqnarray}
       \Fdc^{1\omega}(t)  &= & \alpha_0 \left(\Vdc ^2 + \frac{\Vac^2}{2} \right)
      + \alpha' A\Vdc\Vac \cos\phi. 
      \label{Fdc2'} 
\end{eqnarray}

The terms with $\wm$ are found in $F_0$ (Eq.~\ref{eq:Fdc2}), $\Fw$ (Eq.~\ref{eq:Fw2}) 
and $\Fww$ (Eq.~\ref{eq:F2w2})
and gathering them yields the following result:
\begin{equation}
\begin{split}
       \label{Fw2'} 
       F_{\wm}^{1\omega} (t)  & =   2\alpha_0 \Vdc\Vac \cos(\wm t + \phi) \\
       & \quad + \alpha' A \left\{ \left(\Vdc ^2 + \frac{\Vac^2}{2}\right) \cos (\wm t)
      + \frac{1}{4} \Vac ^2 \cos (\wm t + 2\phi)\right\} \\
    & =  2\alpha_0 \Vdc\Vac \{\cos(\wm t) \cos\phi - \sin(\wm t)\sin\phi\} \\
    &  \quad + \alpha' A \left(\Vdc ^2 + \frac{\Vac^2}{2}\right) \cos (\wm t) \\
    &  \quad +  \frac{1}{4} \alpha' A \Vac ^2 \{\cos (\wm t) \cos(2\phi) - \sin(\wm t)\sin(2 \phi)\}\\
    & =  \left\{\alpha' A \Vdc ^2 + 2\alpha_0 \Vdc\Vac  \cos\phi 
    + \frac{\alpha'A}{2}\left(1 + \frac{1}{2}\cos(2\phi)\right)\Vac^2 \right\}\cos(\wm t) \\
    &  \quad - 2\Vac \left \{\alpha_0 \Vdc \sin\phi+ \frac{1}{4}   \alpha' A \Vac \sin(2\phi) \right\} \sin(\wm t)\\
    & =  \Fin^{1\omega} \cos (\wm t) + \Fout^{1\omega} \sin(\wm t)
  \end{split}
\end{equation}

As we have seen in Sect.~\ref{sec:Effect_of_coherent_force},
the amplitude of $\cos(\wm t)$ and $\sin(\wm t)$ terms, $\Fin^{1\omega}$ and $\Fout^{1\omega}$,
determine the frequency shift, $\df$, and dissipation, $g$, respectively.
Here the superscript, $1\omega$, refers to the case of $\wel = \wm$ and
thus $1\omega$D-KPFM.

$\Fin^{1\omega}$ and $\Fout^{1\omega}$ are expressed as follows:
\begin{equation}
\begin{split}
  \label{eq:Fin_1f}
  \Fin^{1\omega} & =  \alpha' A \Vdc ^2 + 2\alpha_0  \Vdc\Vac  \cos\phi 
    + \frac{\alpha'A}{2}\left(1 + \frac{\cos(2\phi)}{2}\right)\Vac^2 \\
    & =  \alpha'A \left(\Vdc + \frac{\alpha_0 \cos\phi}{ \alpha' A} \Vac \right)^2
    - \left[\frac{\alpha_0^2 \cos^2\phi}{\alpha'A} 
      - \frac{\alpha'A}{2}\left(1+ \frac{\cos(2\phi)}{2}\right)\right]\Vac^2
\end{split}
\end{equation}

\begin{equation}
\begin{split}
  \label{eq:Fout_1f}
  \Fout^{1\omega} & = -  2\Vac \left \{\alpha_0 \Vdc \sin\phi+ \frac{1}{4}   \alpha' A \Vac \sin(2\phi) \right\} \\
  & = -2\alpha_0 \Vac\sin \phi \left(\Vdc + \frac{1}{2}\frac{\alpha' A \cos\phi}{\alpha_0}\Vac \right)
  \end{split}
\end{equation}
Eq.~\ref{eq:Fin_1f} shows that 
$\df$ - $\Vbias $ curve is a parabola whose minima is located at
by $\displaystyle \Vbias = \Vcpd -\frac{\alpha_0 }{ \alpha' A} \Vac \cos\phi$
and $g$ - $\Vbias$ curve is a straight line which intersects $\Fout = 0$ line 
at $\displaystyle \Vbias = \Vcpd  -\frac{1}{2}\frac{\alpha' A }{\alpha_0}\Vac \cos\phi$.

When $\phi=90$ is chosen, we find
\begin{equation}
  \label{eq:Fin_1f_90}
  \Fin^{1\omega} = \alpha'A \left(\Vdc^2+\frac{\Vac^2}{4}\right)
\end{equation}

\begin{equation}
  \label{eq:Fout_1f_90}
  \Fout^{1\omega} = -2\alpha_0 (\Vbias - \Vcpd)\Vac
\end{equation}

Substituting Eq.~\ref{eq:Fin_1f_90} and \ref{eq:Fout_1f_90} into 
Eq.~\ref{eq:df_Fin} and \ref{eq:g_Fout} yields
\begin{align}
  \Delta f & = -\frac{1}{2}\frac{f_0}{k}\frac{\Fin}{A}
  = -\frac{1}{2}\frac{f_0}{k}\alpha' \left\{(\Vbias - \Vcpd)^2 + \frac{\Vac^2}{4}\right\}\\
  g & = g_0 \left(1 - \frac{Q}{kA}\Fout\right)
   = g_0\left\{1 + 2\frac{Q}{kA}\alpha_0(\Vbias - \Vcpd)\Vac\right\}
      \label{eq:g}
\end{align}
 In this case, the dissipation, $g$, is proportional to $\Vdc=\Vbias-\Vcpd$.
It is therefore possible to use the dissipation signal, $g$, 
as the KPFM bias voltage feedback signal
with $g_0$ as its control setpoint value.
It is important to notice that $\df$ is proportional 
to electrostatic force gradient
whereas $g$ is proportional to \textit{electrostatic force}.

\paragraph{Case II , $\wel= 2\wm$: $2\omega$D-KPFM}
\label{sec:2fDKPFM}
In this case, $\wel + \wm = 3 \wm$ and $\wel - \wm = \wm$.
The actual dc term, $\Fdc$, becomes:
\begin{equation}
  \label{eq:Fdc_2f}
         \Fdc^{2\omega}(t) =  \alpha_0 \left(\Vdc ^2 + \frac{\Vac^2}{2} \right)
\end{equation}

The terms with $\wm$ are found in $F_0$ (Eq.~\ref{eq:Fdc2}), $\Fw$ (Eq.~\ref{eq:Fw2}) in this case
and results in
\begin{equation}
  \begin{split}
       F_{\wm}^{2\omega}(t) & = \alpha' A \left\{\left(\Vdc ^2 + \frac{\Vac^2}{2} \right) \cos (\wm t) 
         +  \Vdc\Vac \cos(\wm t + \phi)\right\}\\
       & = \left\{\Vdc^2 + \Vdc\Vac \cos\phi + \frac{\Vac^2}{2}\right\}\cos (\wm t)\\
       & \quad -\alpha' A \Vdc\Vac \sin\phi \sin(\wm t)\\
       & = \Fin^{2\omega} \cos(\wm t) + \Fout^{2\omega} \sin(\wm t)
       \end{split}
\end{equation}
where 
\begin{eqnarray}
  & \Fin^{2\omega} & =  \alpha'A \left(\Vdc + \frac{\Vac}{2} \cos\phi   \right)^2
  + \frac{\alpha'A }{2}\left(1 - \frac{\cos^2\phi}{2} \right) \Vac^2 \label{eq:Fin_2f}\\
 & \Fout^{2\omega} & = - \alpha'A  \Vdc\Vac  \sin\phi \label{eq:Fquad_2f}
\end{eqnarray}

When $\phi=90^\circ$, 
\begin{eqnarray}
  & \Fin^{2\omega} & =  \alpha'A \left(\Vdc^2 + \frac{\Vac^2}{2}\right)
                     \label{eq:Fin_2f_90}\\
 & \Fout^{2\omega} & = - \alpha'A  \Vdc\Vac  \label{eq:Fquad_2f_90}
\end{eqnarray}

In contrast to $\wel = \wm$ case (Eq.~\ref{eq:Fout_1f}), 
$\Fout$ is proportional to $\alpha'$ rather than $\alpha$,
indicating $g$ is sensitive to \textit{electrostatic force gradient}.
Substituting Eqs.~\ref{eq:Fin_2f_90} and \ref{eq:Fquad_2f_90} into 
Eqs.~\ref{eq:df_Fin} and \ref{eq:g_Fout} yields:
\begin{align}
  \Delta f & = -\frac{1}{2}\frac{f_0}{k}\frac{\Fin}{A}
  = -\frac{1}{2}\frac{f_0}{k}\alpha' \left\{(\Vbias - \Vcpd)^2 + \frac{\Vac^2}{2}\right\}\\
  g & = g_0 \left(1 - \frac{Q}{kA}\Fout\right)
   = g_0\left\{1 + \frac{Q}{k}\alpha'(\Vbias - \Vcpd)\Vac\right\}.
      \label{eq:g}
\end{align}

Here we notice the following two important points:
(1) the apparent shift of $\Vcpd$ appearing in  $\df$-$\Vbias$ curve
does not depend on $\alpha$ and is just determined by $\phi$ and $\Vac$,
(2) there is no apparent shift of $\Vcpd$ in a $g$-$\Vbias$ curve.
The point (2) indicates an important advantage of $2\omega$D-KPFM 
over $1\omega$D-KPFM as there is no need 
for carefully adjusting the phase, $\phi$,
for accurate $\Vcpd$ measurements.

\section{Experimental}
\label{sec:Experimental}
\begin{figure}[t]
  \centering
  \includegraphics[width=90mm]{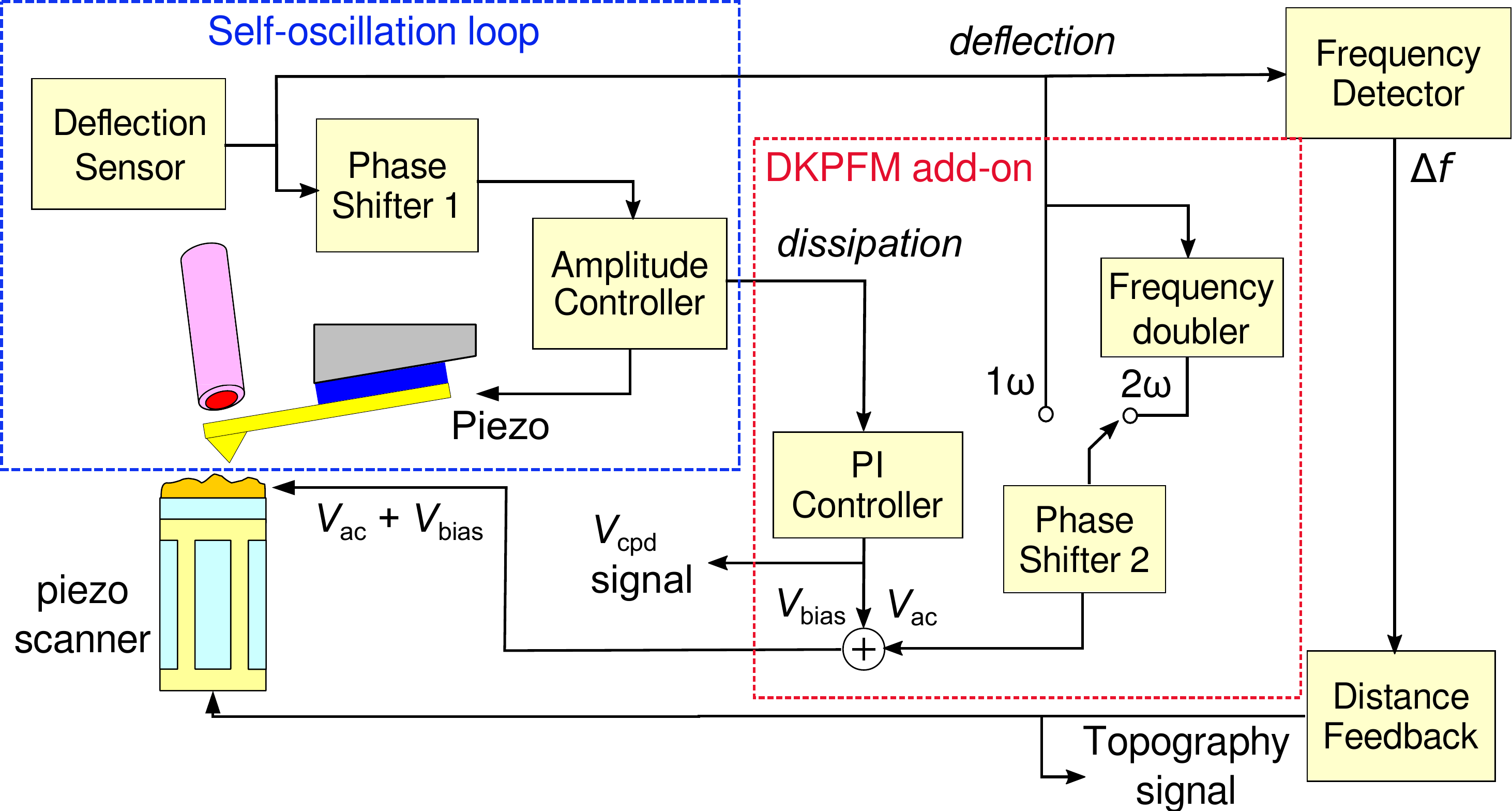}
  \caption{Block diagram of the experimental setup for force sensitive ($1\omega$D-KPFM) 
    and force-gradient sensitive ($2\omega$D-KPFM) dissipation modulated KPFM techniques.
    Reprinted from~\cite{Miyahara2017b}, with the permission of AIP Publishing
    \label{fig:DKPFMDiagram2}}
\end{figure}
Figure~\ref{fig:DKPFMDiagram2} depicts the block diagram of the experimental 
setup used for both $1\omega$ and $2\omega$D-KPFM measurements.
The setup is based on the self-oscillation mode FM-AFM system~\footnote{D-KPFM will
also work with the phase-locked loop tracking oscillator method
in which the cantilever oscillation is excited 
by an external oscillator~\cite{Durig1997}.}~\cite{Albrecht1991}.
While two additional components, a phase shifter,
a proportional-integrator (PI) controller, are required 
for both $1\omega$ and $2\omega$D-KPFM operation,
a frequency doubler that generates an ac voltage with the frequency of $2\wm$
from the cantilever deflection signal
is also required for $2\omega$D-KPFM operation.
The amplitude controller that consists of a root-mean-square (RMS) 
amplitude detector and a PI controller (NanoSurf easyPLLplus oscillator controller)
is used to keep the amplitude of tip oscillation constant.
The detection bandwidth of the RMS amplitude detector 
is extended to about 3~kHz by replacing the integration capacitor in the original
RMS detector circuit.
The output of the amplitude controller is the dissipation signal.

The cantilever deflection signal is fed into the additional phase shifter 
(Phase shifter 2),
which serves to adjust the relative phase, $\phi$, 
to produce the coherent ac voltage that is 90$^\circ$ out of phase 
with respect to the cantilever deflection oscillation.
The frequency doubler is used to produce a sinusoidal ac voltage with 
two times the tip oscillation frequency.
The dissipation signal acts as the input signal to the PI controller,
which adjusts $\Vbias$ 
to maintain a constant dissipation 
equal to the value without $\Vac$ applied, $g_0$.
Although in the following experiments 
we used a digital lock-in amplifier (HF2LI, Zurich Instruments)
operated in the external reference mode as a phase shifter 
as well as a frequency doubler for convenience,
other simpler phase shifter circuits such an all-pass filter
can also be used for the D-KPFM measurements.

We used a JEOL JSPM-5200 atomic force microscope for the experiments 
with the modifications described below.
The original laser diode was replaced by a fiber-optic collimator 
with a focusing lens that is connected 
to a fiber-coupled laser diode module (OZ Optics).
The laser diode was mounted on a temperature controlled fixture 
and its driving current was modulated with a radio frequency signal 
with a RF bias-T (Mini-Circuits: PBTC-1GW).
to reduce the deflection detection noise~\cite{Fukuma05}.
The deflection noise density floor as low as 13\,fm$/\sqrt{\text{Hz}}$ was achieved.
An open source scanning probe microscopy control software GXSM
was used for the control and data acquisition~\cite{Zahl2010}.
A commercial silicon AFM cantilever (NSC15, MikroMasch) 
with a typical spring constant of about 20\,N/m and resonance frequency 
of $\sim 300$\,kHz was used in high-vacuum environment with
the pressure of $1 \times 10^{-7}$\,mbar.
The ac and dc voltages was applied to the sample with respect 
to the grounded tip to minimize the capacitive crosstalk~\cite{Diesinger2008}
 to the piezoelectric
plate used for driving cantilever oscillation
that is located beneath the cantilever support.

\section{Results and discussion}
\subsection{Validation of  D-KPFM theory}
\label{sec:validation}
In order to validate the analysis described in the previous sections
~\ref{sec:1fDKPFM} and \ref{sec:2fDKPFM}, 
$\df$-$\Vbias$ and $g$-$\Vbias$ curves are measured with the coherent ac
voltage with different phases applied.

\paragraph{$1\omega$D-KPFM case}
\begin{figure}[t]
  \centering
   \includegraphics[width=80mm]{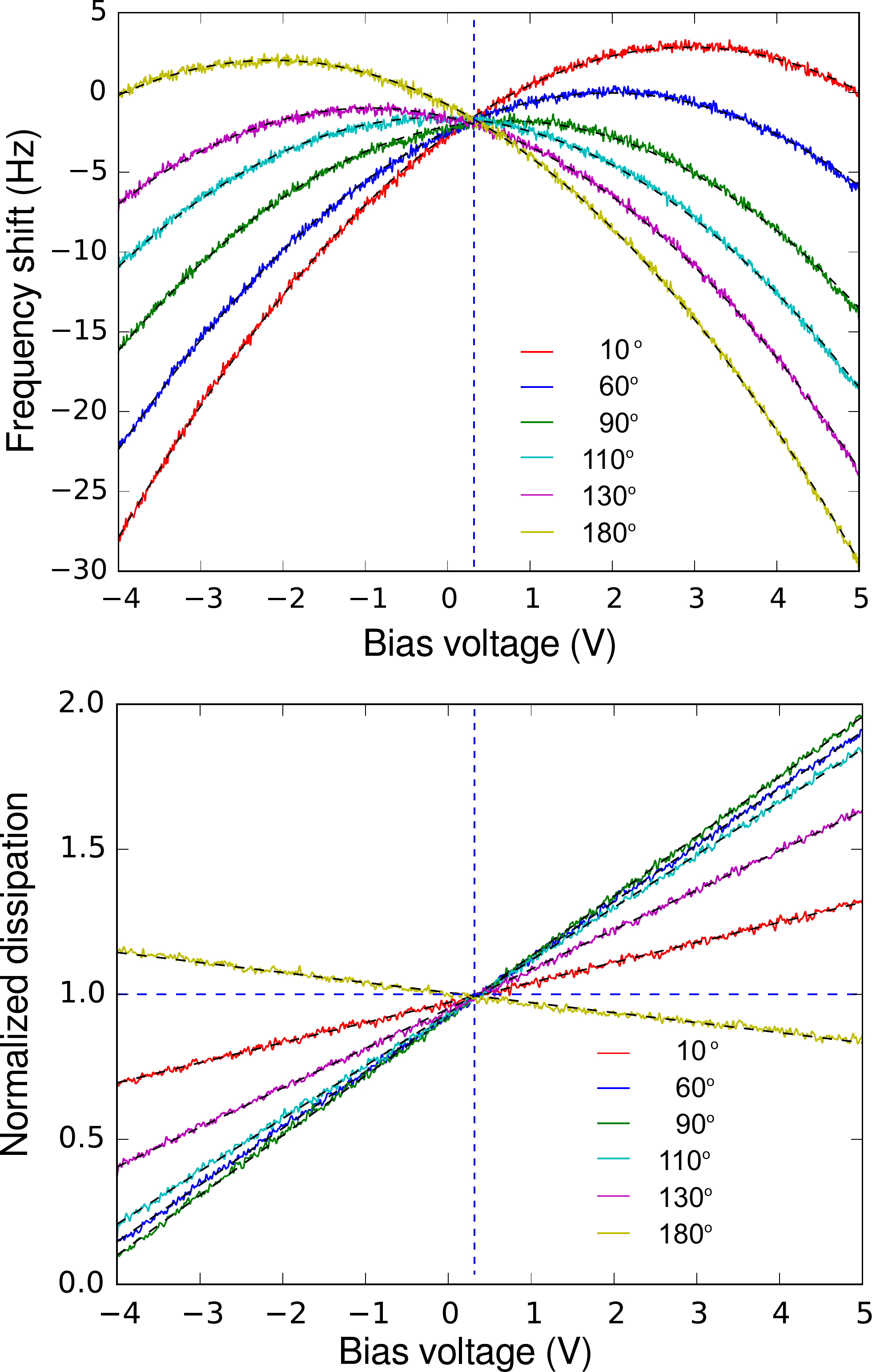}   
  \caption{Validation of $1\omega$D-KPFM theory. 
    (a) Frequency shift, $\df$, and (b) dissipation signal, $g$, 
  versus dc bias voltage, $\Vbias$, curves 
  taken with a coherent sinusoidally oscillating voltage 
  with the amplitude, $\Vac=100$\,mV$_\mathrm{p-p}$ and various phases, $\phi$,
  applied to a 200\,nm thick SiO$_2$ on Si substrate.
  The dissipation signal is normalized with the value without 
  the ac bias voltage (indicated with the horizontal blue dashed line).
  In both figures, each of dashed lines represent fitted curves 
  assuming a parabola for $\df$ and a linear line for $g$ as indicated
  in Eq.~\ref{eq:Fin_1f} and \ref{eq:Fout_1f}, respectively.
  The oscillation amplitude of the tip was 7.2\,nm$_\mathrm{p-p}$ 
  and the quality factor of the cantilever was 9046.
  Reprinted with permission from~\cite{Miyahara2015}.
  Copyright 2016 by the American Physical Society
  \label{fig:df-V_1f}}. 
\end{figure}

Figure~\ref{fig:df-V_1f} shows simultaneously measured 
$\df$ and $g$ versus $\Vbias$ curves 
with a coherent sinusoidally oscillating voltage 
with the amplitude, $\Vac=100$\,mV$_\mathrm{p-p}$ and various phase, $\phi$.
The curves were taken on a Si substrate with 200~nm thick oxide SiO$_2$.
A fitted curve with a parabola for $\df$-$\Vbias$ curves
(Eq.~\ref{eq:Fin_1f})
or a linear line for $g$-$\Vbias$ curves (Eq.~\ref{eq:Fout_1f}) 
is overlaid on each experimental curve,
indicating a very good agreement between the theory and experiments.
As can be seen in Fig.~\ref{fig:df-V_1f}(a) and (b), 
the position of the parabola vertex shifts 
and the slope of $g$-$\Vbias$ curve changes 
systematically with varying phase. 

\begin{figure}[t]
  \centering
  \includegraphics[width=85mm]{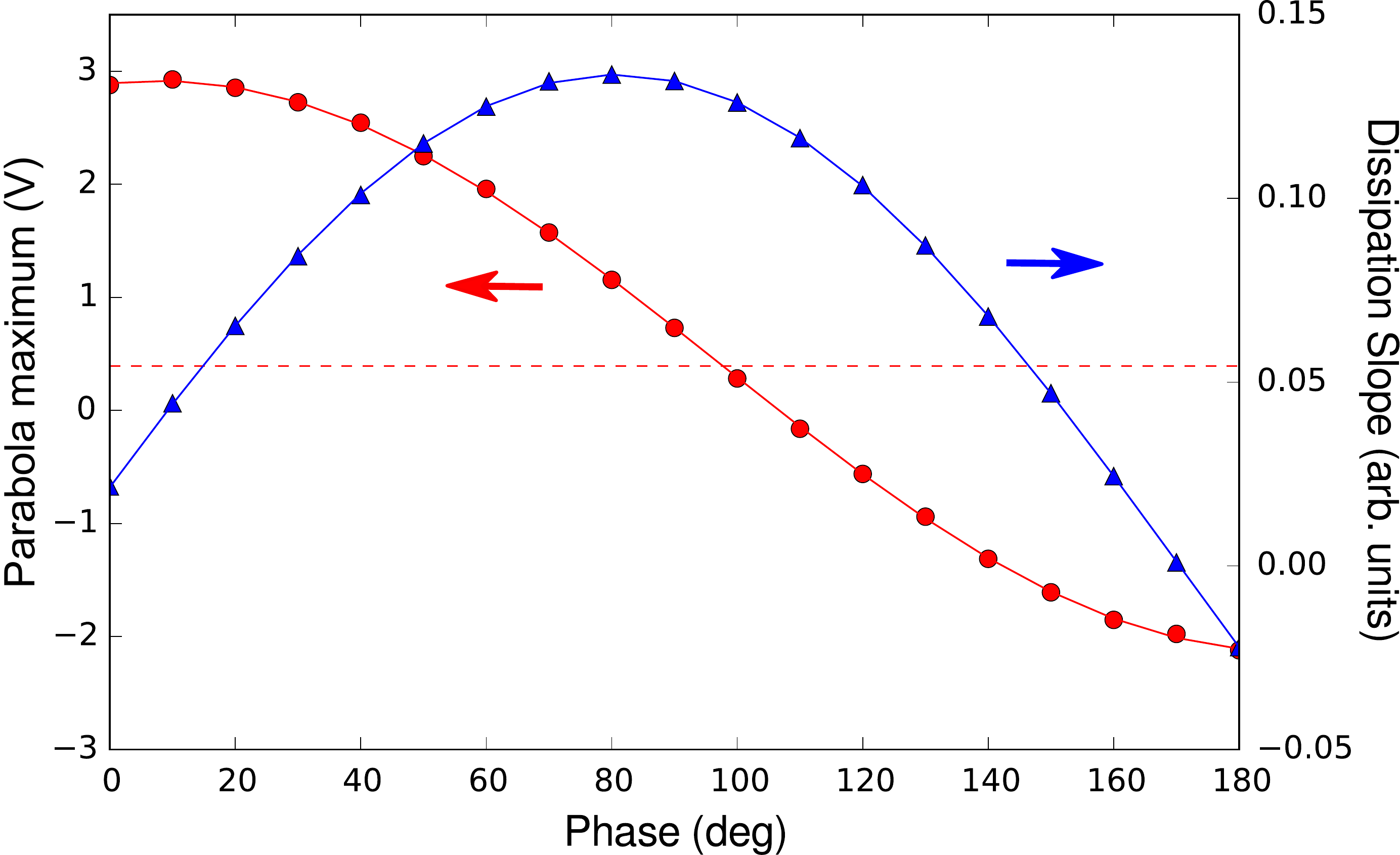}   
  \caption{Parabola maximum of the measured $\df$-$\Vbias$ curves 
  (red circles) (Fig.~\ref{fig:df-V_1f}(a))
  and the dissipation slope of $g$-$\Vbias$ curves (blue circles)
  (Fig.~\ref{fig:df-V_1f}(b)). 
  Each solid line represents the fitted curve with the cosine function (Eq.~\ref{eq:Fin_1f})
  for the parabola minimum and with the sine function (Eq.~\ref{eq:Fout_1f})
  for the dissipation slope.
  The horizontal dashed line indicates the voltage for parabola maximum without 
  the ac bias voltage.
  Reprinted with permission from~\cite{Miyahara2015}.
  Copyright 2016 by the American Physical Society
  \label{fig:fit_parameters_1f}}
\end{figure}

In order to further validate the theoretical analysis, 
the voltage coordinate for the vertices (parabola maximum) of $\df$-$\Vbias$ curves 
and the slope of $g$-$\Vbias$ lines (dissipation slope) are plotted 
against the phase, $\phi$, in Fig.~\ref{fig:fit_parameters_1f}.
Each plot is overlaid with a fitted curve (solid curve) with the cosine function
(Eq.~\ref{eq:Fin_1f})
for the parabola maximum and with the sine function (Eq.~\ref{eq:Fout_1f}) 
for the dissipation slope,
demonstrating an excellent agreement between the experiment and theory.
The parabola maximum versus phase curve intersects 
that for the parabola without ac voltage at the phase of 97$^\circ$ 
as opposed to 90$^\circ$ which is predicted by the theory.
This deviation is mainly due to the phase delay 
in the photodiode preamplifer electronics.
The dissipation slope takes its maximum value at around 81$^\circ$, 
again deviating from the theoretical value of 90$^\circ$.
This deviation is probably due to the residual capacitive crosstalk 
to the excitation piezo~\cite{Diesinger2008, Melin2011}.

\paragraph{$2\omega$D-KPFM case}
\begin{figure}[t]
  \centering
  \includegraphics[width=85mm]{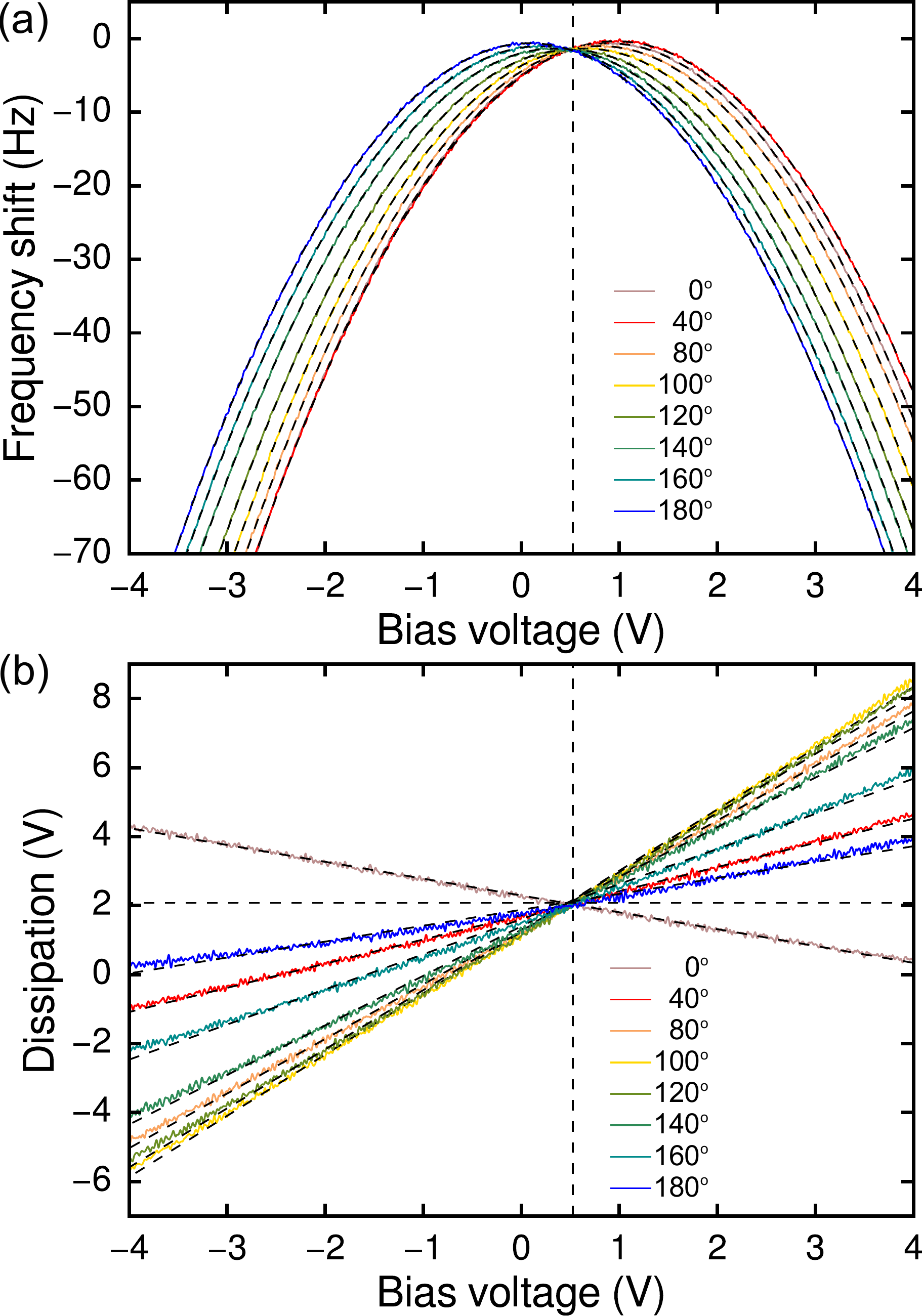}
 \caption{(a) Frequency shift, $\df$, and (b) dissipation signal, $g$, 
  versus dc bias voltage, $\Vbias$, curves 
  taken with a coherent sinusoidally oscillating voltage 
  with $\wel=2\wm$, $\Vac=1$~V 
  and various $\phi$,
  applied to a template stripped gold substrate.
  The vertical dashed line in (a) and (b) indicates $\Vcpd$ which is measured 
  as the voltage coordinate of the parabola vertex without $\Vac$.
  The horizontal dashed line in (b) indicates the dissipation without $\Vac$.
  In both figures, each of dashed lines represent fitted curves 
  assuming a parabola for $\df$ and a linear line for $g$ as indicated
  in Eqs.~\ref{eq:Fin_2f} and \ref{eq:Fquad_2f}, respectively.
  The oscillation amplitude of the tip was 10~nm$_\mathrm{p-p}$ 
  and the quality factor of the cantilever was 25000
\label{fig:df-V_2f}}
\end{figure}

In order to validate Eqs.~\ref{eq:Fin_2f} and \ref{eq:Fquad_2f},
$\df$-$\Vbias$ and $g$-$\Vbias$ curves were measured 
while a coherent sinusoidally oscillating voltage 
with $\wel = 2\wm$, $\Vac=1$~V (2~V$_\mathrm{p-p})$
and various phases, $\phi$,
is superposed with $\Vbias$.

\begin{figure}[t]
  \includegraphics[width=120mm]{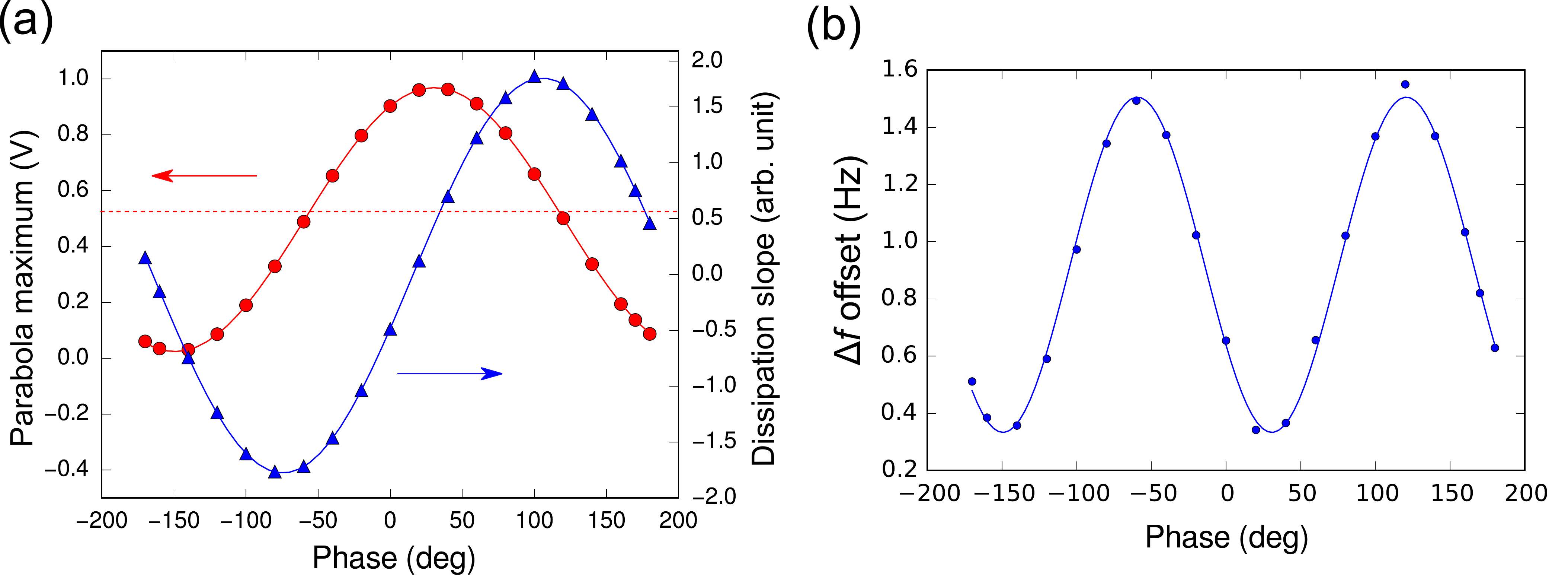}   
  \caption{(a) Voltage coordinate of the vertices of the measured 
    $\df$-$\Vbias$ curves (parabola maximum) 
  (red circles) extracted from the results Fig.~\ref{fig:df-V_2f}(a)
  and the slope of dissipation-$\Vbias$ curves (blue triangles) extracted from
  Fig.~\ref{fig:df-V_2f}(b) as a function of phase. 
  Each solid line represents the fitted curve with the cosine function (Eq.~\ref{eq:Fin_2f})
  for the parabola maximum and with the sine function (Eq.~\ref{eq:Fquad_2f})
  for the dissipation slope.
  The horizontal dashed line indicates the voltage coordinate of the parabola 
  measured without the ac bias voltage.
  (b) $\df$ offset as a function of phase. 
  The solid line represents the fitted curve with 
  the second term of Eq.~\ref{eq:Fin_2f}.
  Reprinted from~\cite{Miyahara2017b}, with the permission of AIP Publishing
  \label{fig:fit_parameters_2f}}
\end{figure}

Figure~\ref{fig:df-V_2f}(a) and (b) show the simultaneously measured 
$\df$ and $g$ versus $\Vbias$ curves, respectively.
The curves are taken on a template stripped gold surface.
A fitted curve with a parabola for each of the $\df$-$\Vbias$ curves
(Eq.~\ref{eq:Fin_2f})
or with a linear line for each of the $g$-$\Vbias$ curves (Eq.~\ref{eq:Fquad_2f}) 
is overlaid on each experimental curve,
indicating a very good agreement between the theory and experiments.
As can be seen in Fig.~\ref{fig:df-V_2f}(a) and (b), 
the position of the parabola vertex shifts 
both in $\Vbias$ and $\df$ axes
and the slope of $g$-$\Vbias$ curve changes 
systematically with the varied phase, $\phi$. 
 For further validating the theory,
the voltage coordinate of the parabola vertex (parabola maximum voltage) 
of each $\df$-$\Vbias$ curve and 
the slope of each $g$-$\Vbias$ curve are plotted against $\phi$
in Fig.~\ref{fig:fit_parameters_2f}.
Each plot is overlaid with a fitted curve (solid curve) with the cosine function
(see Eq.~\ref{eq:Fin_2f})
for the parabola minimum voltage 
and with the sine function (Eq.~\ref{eq:Fquad_2f}) 
for the dissipation slope,
demonstrating an excellent agreement between the experiment and theory.
The $\Vbias$ dependence of the frequency shift coordinate 
of the parabola vertices 
(frequency shift offset) also shows a very good agreement with the theory 
(second term of Eq.~\ref{eq:Fin_2f}) as shown in Fig.~\ref{fig:fit_parameters_2f}(b).
The parabola minimum voltage versus phase curve intersects 
that of $\df$-$\Vbias$ without ac bias voltage at $\phi = 121^\circ$.
The deviation from the theoretically predicted value of 90$^\circ$ 
is due to the phase delay in the detection electronics.
We also notice that the amplitude of parabola minimum versus phase curve is
0.472~V which is in good agreement with 0.5~V predicted by the theory
($\Vac /2$ in Eq.~\ref{eq:Fin_2f}).

\subsection{Illustrative example of D-KPFM imaging}
\begin{figure}[t]
  \centering
  \includegraphics[width=120mm]{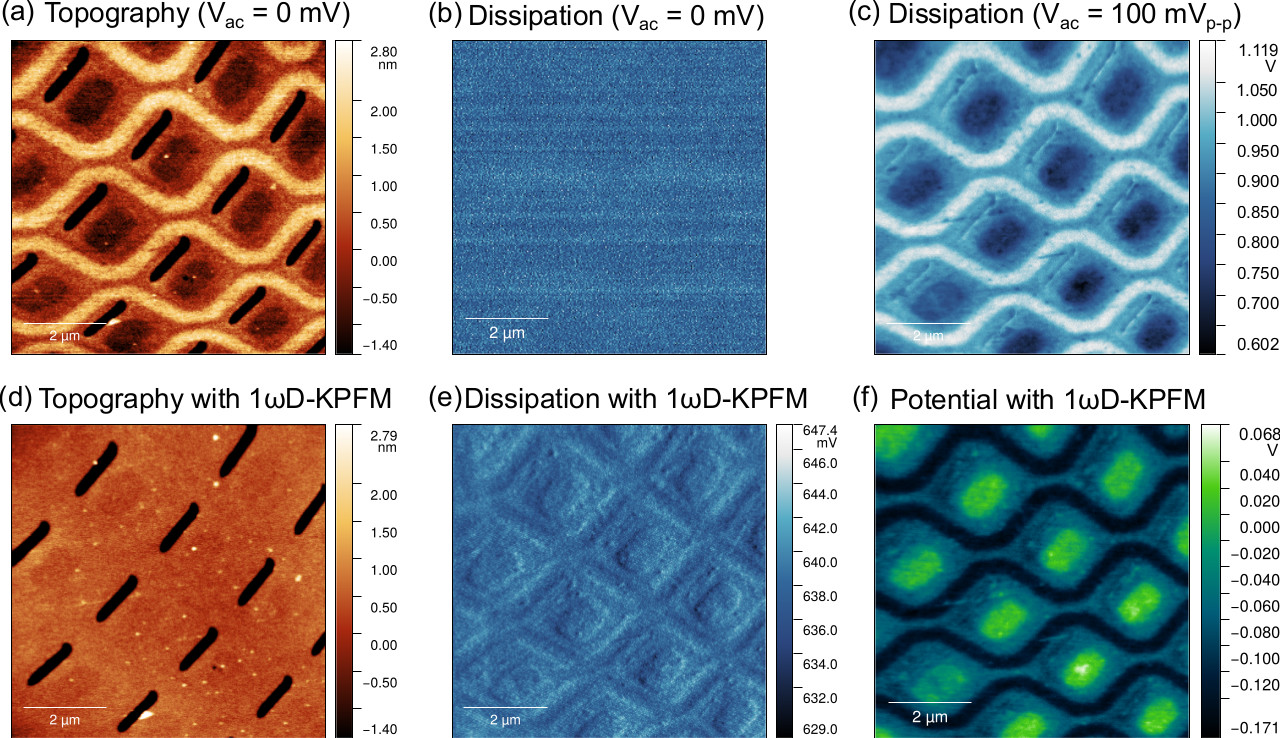}   
  \caption{(a) Topography and (b) dissipation images of a CCD image sensor 
  with FM mode with no ac voltage 
   and a constant $\Vbias = 50$~mV applied. $\df = -0.3$~Hz. 
   No contrast is observed in (b).
  (c) dissipation image with coherent ac voltage with FM mode with its amplitude, 
  $\Vac=100$~mV$_\mathrm{p-p}$ ($\wel = \wm$).
  The similar contrast to that appears in the topography (a) is observed.
  (e) topography, (f) dissipation, (g) potential images 
  with $1\omega$D-KPFM technique with $\Vac=100$~mV$_\mathrm{p-p}$, $\wel=\wm$.
  \label{fig:demo_DKPFM}}
\end{figure}

Fig.~\ref{fig:demo_DKPFM} shows an example to illustrate
how this technique works.
Fig.~\ref{fig:demo_DKPFM}(a) and (b) show the topography and dissipation images
of a CCD image sensor taken with FM mode without applying ac voltage.
at a constant dc voltage of 50~mV.
The topography contrast in Fig.~\ref{fig:demo_DKPFM}(a) is 
a convolution of the true topography and the electrostatic contrast 
except for black slits which are actual recesses.
This 'apparent topography' contrast is caused by the different electrostatic force 
on the differently doped regions (bright: n-doped, dark: p-doped)~\cite{Sadewasser03, Miyahara2012}.
The dissipation image with no ac voltage applied (Fig.~\ref{fig:demo_DKPFM}(b)) 
shows no contrast,
indicating no observable Joule dissipation
due to the mismatch of the tip oscillation frequency and the dielectric relaxation time
of the sample under these imaging condition~\cite{Denk91}.
It is important to notice that the dissipation signal in FM-AFM is sensitive only 
to the tip-sample interaction force 
with the time (phase) delay comparable to the tip oscillation period as demonstrated 
in the previous sections (Sect.~\ref{sec:validation}).
Fig.~\ref{fig:demo_DKPFM}(c) shows the very similar contrast to the apparent topography (a).
Fig.~\ref{fig:demo_DKPFM}(d), (e) and (f) shows the simultaneously taken topography, 
dissipation and CPD image with $1\omega$D-KPFM technique. 
While the apparent topography in (a) disappears in (d),
the similar contrast instead appears in the CPD image (f).
Here, the dissipation image (e) is now the input for the KPFM bias feedback controller 
(\textit{e.g.} error signal for PI controller in Fig.~\ref{fig:DKPFMDiagram2})
and the CPD image is the output of the PI controller.
The electrostatic contrast which appears as the apparent topography contrast 
is thus transferred to the CPD contrast through the induced dissipation.

\subsection{Comparison of different KPFM techniques}
\begin{figure}[t]
  \centering
  \includegraphics[width=60mm]{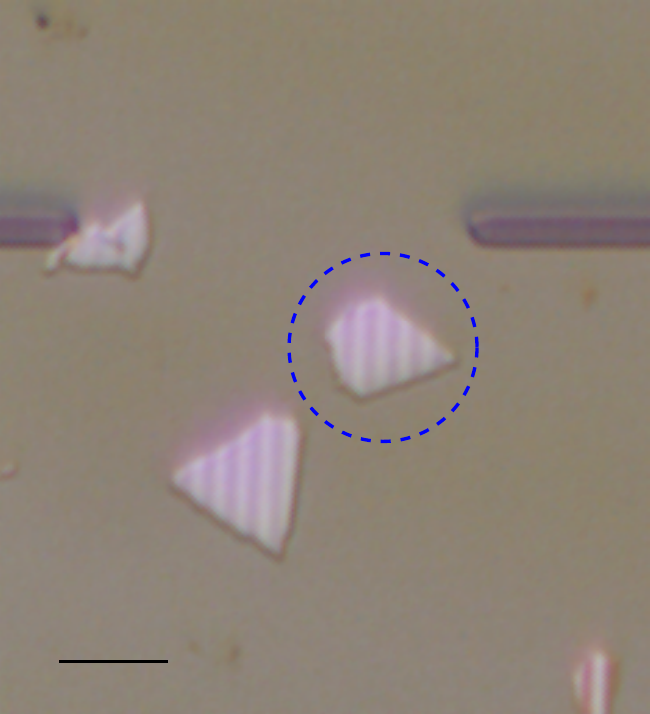}   
  \caption{Optical micrograph of the patterned MoS$_2$ flakes exfoliated on 
    SiO$_2$/Si substrate.The flake in the circle is the one imaged by the KPFM imaging
    experiments. Scale bar is 20~$\mu$m.
  \label{fig:optical_sample}}
\end{figure}

In this section, we compare the KPFM images taken with FM-KPFM, AM-KPFM, $1\omega$D-KPFM
and $2\omega$D-KPFM to evaluate the performance of each technique.
We used a patterned MoS$_2$ flake for the comparison.
The several ten $\mu$m scale size of the sample 
together with the etched stripe pattern with 2~$\mu$m pitch 
enables to find the same region of the sample even in the separate runs.
Flakes of MoS$_2$ were deposited onto a SiO$_2$/Si substrate 
by mechanical exfoliation
and a stripe pattern was created by electron beam lithography 
and the subsequent reactive ion etching on top of the flakes.
Fig.~\ref{fig:optical_sample} shows the optical micrograph of the MoS$_2$ flake
used for the KPFM imaging experiments.
\begin{figure}[hhh]
  \centering
  \includegraphics[width=90mm]{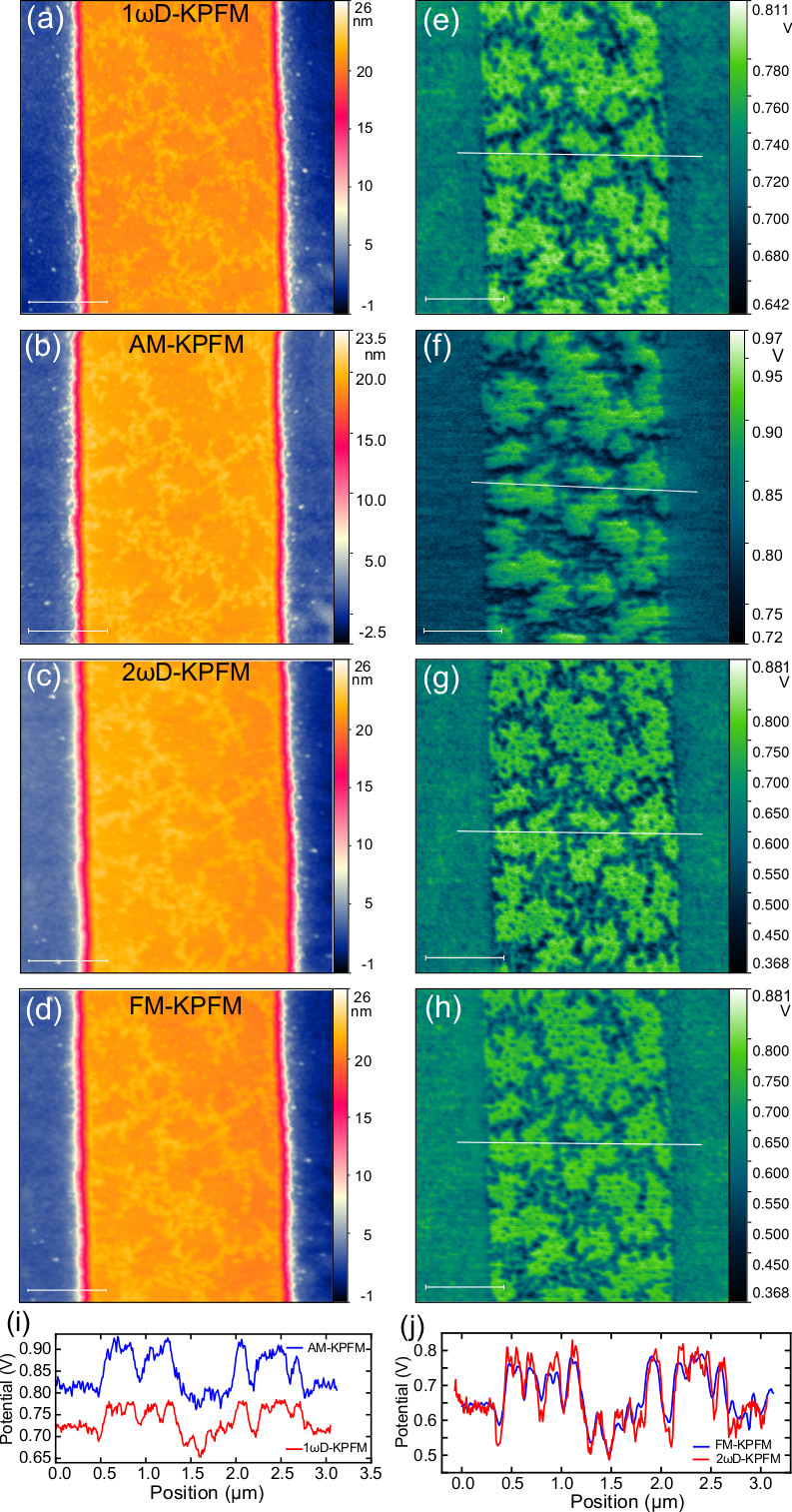}   
  \caption{Topography and CPD images of the patterned MoS$_2$ flake.
   Topography images taken by (a)~$1\omega$D-KPFM, (b)~AM-KPFM, 
   (c)~$2\omega$D-KPFM and (d)~FM-KPFM techniques.
   CPD images taken by (e)~$1\omega$D-KPFM, (f)~AM-KPFM, 
   (g)~$2\omega$D-KPFM and (h)~FM-KPFM techniques.
   (i) Profiles of the CPD images by (e)~$1\omega$D-KPFM and (f)~AM-KPFM techniques.
   (j) Profiles of the CPD images by (g)~$2\omega$D-KPFM and (h)~FM-KPFM techniques
    \label{fig:comparison_KPFM}}
\end{figure}

Fig.~\ref{fig:comparison_KPFM} shows topography and potential images 
of the patterned MoS$_2$ flake
by (a)~$1\omega$D-KPFM, (b)~AM-KPFM, (c)~$2\omega$D-KPFM and (d)~FM-KPFM techniques.
The $1\omega$D-KPFM, $2\omega$D-KPFM and FM-KPFM images 
were taken with the same cantilever tip 
($f_0 = 306553$\,Hz, $k=20.8$\,N/m, $Q=14963$) and
the AM-KPFM image was taken with a different one
($f_0 = 298044$\,Hz, $k=27.2$\,N/m, $Q=14700$). 
These two cantilevers were of the same type  (NSC15, MikroMasch) and 
taken from the same batch.
In $1\omega$D-KPFM imaging, a sinusoidally oscillating voltage 
with an amplitude of $\Vac=100$\,mV$_\mathrm{p-p}$ 
phase-locked with the tip oscillation was applied to the sample.
In AM-KPFM imaging, a sinusoidally oscillating voltage 
with an amplitude of $\Vac=8$\,V$_\mathrm{p-p}$ 
whose frequency was tuned to the second flexural resonance peak 
(resonance frequency:1,903,500\,Hz, quality factor:2400)
was applied to the sample.
The resulting oscillation amplitude was detected by a high-speed
lock-in amplifier (HF2LI, Zurich Instruments)
and used for the dc bias voltage feedback~\cite{Kikukawa96, Sommerhalter1999}.
In FM-KPFM imaging, a sinusoidally oscillating voltage 
with the amplitude of $\Vac=2.0$\,V$_\mathrm{p-p}$
and frequency of 300\,Hz was applied to the sample~\cite{Kitamura98}.
The number of pixels in the images is 512$\times$512.
The scanning time for $1\omega$D-KPFM, $2\omega$D-KPFM and FM-KPFM imaging was 
512\,s/frame (1\,s/line).
The scanning time for AM-KPFM was for 1024\,s/frame (2\,s/line).
In all the imaging modes, the frequency shift signal was used for
topography feedback.

The topography images (Fig.~\ref{fig:comparison_KPFM}(a-d))  
show an unetched terrace located between the etched regions. 
The height of the terrace is approximately 20~nm 
with respect to the etched regions.
A clear fractal-like pattern can be seen on the terrace in all the CPD images.
The potential contrast can be ascribed to the residue of the etch resist (PMMA)
as the topography images show a similar contrast with a thickness of about 1\,nm.

\paragraph{Comparison between $1\omega$D-KPFM and AM-KPFM}
First, we compare the CPD images taken 
by $1\omega$D-KPFM (Fig.~\ref{fig:comparison_KPFM}(a))
and AM-KPFM (Fig.~\ref{fig:comparison_KPFM}(b)). 
The CPD image by $1\omega$D-KPFM shows better clarity than that by AM-KPFM.
The difference is due to the lower signal-to-noise ratio of the amplitude signal
of the second flexural mode oscillation 
despite that a much higher $\Vac=8$\,V$_\mathrm{p-p}$ was applied in AM-KPFM.
The higher effective spring constant of the second flexural mode oscillation
($k_\mathrm{2nd} \sim 40 \times k \approx 800$\,N/m)
compared to the first mode~\cite{Kikukawa96, Melcher07} 
and lower observed quality factor ($Q_\mathrm{2nd} = 2400$ 
compared with 14700 for the first mode)
account for the difference as the signal in both operating modes is proportional to $Q/k$.
Notice that the cantilevers with lower effective spring constant 
($k \approx 2 \sim 3 $\,N/m) have been used in the most reported AM-KPFM measurements 
in which $\Vac$ of the order of 100\,mV is employed 
\cite{Kikukawa96, Sommerhalter1999, Glatzel2003a, Zerweck05, Burke09a}. 
Fig.~\ref{fig:comparison_KPFM}(i) shows the line profiles of potential on the same
location indicated as a white line in the CPD images 
Fig.~\ref{fig:comparison_KPFM}(e) and (f).
Two profiles are in a very good agreement except the constant offset.
This observed similarity can be understood by the fact that 
both $1\omega$D-KPFM and AM-KPFM are sensitive to the electrostatic force.
The constant offset between two profiles is probably due to the different tips 
used in the two separate experiments.

\paragraph{Comparison between $2\omega$D-KPFM and FM-KPFM}
Now we turn to the comparison between $2\omega$D-KPFM 
(Fig.~\ref{fig:comparison_KPFM}(g)) and FM-KPFM (Fig.~\ref{fig:comparison_KPFM}(h)).
The CPD images taken with $2\omega$D-KPFM and FM-KPFM look very similar
and the line profiles of two images taken on the same location 
indicated as a white line show a good agreement.
A closer inspection reveals slightly larger potential contrast 
in the $2\omega$D-KPFM image compared with the FM-KPFM image,
which is due to the faster KPFM feedback response of $2\omega$D-KPFM
by virtue of the absence of low frequency modulation which is required 
for FM-AFM.

\paragraph{Comparison between $1\omega$D-KPFM and $2\omega$D-KPFM }
Although both potential images taken with $1\omega$D-KPFM (Fig.~\ref{fig:comparison_KPFM}(e))
and $2\omega$D-KPFM (Fig.~\ref{fig:comparison_KPFM}(g))
show the almost identical pattern on the terrace with the nearly same signal-to-noise ratio,
we notice lower contrast in the CPD image taken 
with $1\omega$D-KPFM than that with $2\omega$D-KPFM
from the inspection of the line profiles, Fig.~\ref{fig:comparison_KPFM}(i) and (j).
The peak-to-peak value of the potential variation in the $1\omega$D-KPFM image 
is $\sim 0.12$\,V,
about one half that in the $2\omega$D-KPFM image ($\sim 0.3$\,V).
The similar difference in the potential contrast taken with FM-KPFM and AM-KPFM 
has also been reported in the literature 
and is ascribed to the fact that
the AM-KPFM is sensitive to electrostatic force 
whereas FM-KPFM uses the modulation in the resonance frequency shift 
which is sensitive to force gradient~\cite{Zerweck05, Glatzel2003a}.
As we have already discussed, the same argument applies to $1\omega$D-KPFM
and $2\omega$D-KPFM.
The smaller potential contrast observed in $1\omega$D-KPFM than in $2\omega$D-KPFM
can be explained by larger spatial average due to the stray capacitance
including the body of the tip and the cantilever 
~\cite{Hochwitz1996,Jacobs1999,Strassburg2005a,Cohen2013}.
We noticed that an ac voltage amplitude (2\,V) much larger than that require 
for $1\omega$D-KPFM (100\,mV) was necessary for $2\omega$D-KPFM.
This indicates $\alpha > \alpha'A$ which is determined by the distance
dependence of the tip-sample capacitance. 
This condition could be changed by engineering the tip shape.

In spite of lower potential contrast, $1\omega$D-KPFM has a clear advantage 
in that it requires much smaller $\Vac= 100\,$mV$_\mathrm{p-p}$ 
compared with 2\,V$_\mathrm{p-p}$ for $2\omega$D-KPFM.
This advantage is important for electrical characterizations
of technically relevant materials whose electrical properties
are influenced by the externally applied electric field 
as is the case in semiconductors
where the influence of the large $\Vac$ can induce band-bending effects.

\subsection{Dynamic response of D-KPFM}
The detection bandwidth of D-KPFM is determined 
by the bandwidth of the amplitude control feedback loop used in FM-KPFM.
In fact, we notice that applying the coherent $\Vac$ 
causing dissipative force can be used 
to measure the dynamics of the amplitude control feedback system.

In FM-KPFM, 
the AFM cantilever serves as the frequency determining element 
of a self-driven oscillator
so that the oscillation frequency keeps track 
of the resonance frequency of the cantilever.
In this way, the conservative force has no influence on the drive amplitude
(dissipation signal)
when the time delay set by the phase shifter is property adjusted 
as already discussed in Sect.~\ref{sec:FM-AFM theory}
whereas the amplitude controller compensates for 
the effective $Q$ factor change caused by dissipative force.
Therefore, by modulating $\Vbias$ at a low frequency ($<$ a few\,kHz)
together with applying the coherent $\Vac$, 
the amplitude of the dissipative force can be modulated 
as can been seen in Eqs.~\ref{eq:Fout_1f_90} and ~\ref{eq:Fquad_2f_90}
and the frequency response of the amplitude feedback loop can thus be measured
by demodulating the dissipation signal with a lock-in amplifier.
The measured $-3$\,dB bandwidth of the amplitude feedback loop 
is as high as 1\,kHz,
which is wider than that of the typical PLL frequency detector 
(400\,Hz in this experiment).

Notice that in contrast to the settling time of the oscillation amplitude
of a cantilever subject to a change in conservative force
which is $\tau \sim Q/f_0$~\cite{Albrecht1991},
the response time of the dissipation signal is not limited by $Q$ 
and can be faster because of the active damping mechanism built in 
the amplitude control feedback loop~\cite{Durig1997}
as well as 
the induced energy dissipation which is given by $\pi \Fout A$
in addition to the internal dissipaiton of the cantilever, $\pi k A^2/Q$.
The active damping behavior of the amplitude controller indeed manifests itself 
in the effective negative dissipation shown in $g$-$\Vbias$ curves
(Figs.~\ref{fig:df-V_1f} and \ref{fig:df-V_2f}).
This explains the observed fast response of the amplitude feedback loop,
resulting in the wider bandwidth of the voltage feedback loop in D-KPFM
than that in FM-KPFM that is limited by PLL demodulation bandwidth
(typically $ <1$\,kHz) which sets the bias modulation frequency.

The low frequency $\Vbias$ modulation can also be used for D-KPFM imaging
in the case where other dissipative interactions 
such as the Joule dissipation~\cite{Denk91} 
and single-electron tunneling~\cite{Cockins2010a,Miyahara2017} 
or the dissipation artifact caused by the crosstalk to the frequency shift~\cite{Labuda2011} 
contribute to the dissipation signal.
It is possible to separate the induced electrostatic dissipation
just as is done in FM-FKPFM with $\df$ 
and in DM-KPFM with the tip oscillation amplitude~\cite{Fukuma2004}. 
The modulated dissipation signal due to $\Vbias$ modulation can be demodulated 
with a lock-in amplifier which can then be used for the KPFM bias feedback.

\section{Conclusion}
We reported D-KPFM, a new experimental technique for KPFM 
in which the dissipation signal of FM-AFM is used for KPFM bias-voltage feedback.
We show that D-KPFM can be operated in two different modes, $1\omega$D-KPFM
and $2\omega$D-KPFM which are sensitive to electrostatic force and 
electrostatic force gradient, respectively.
The technique features the simpler implementation and faster scanning than
the most commonly used FM-KPFM technique
as it requires no low frequency modulation.

We provided the theory of D-KPFM by combining two key aspects:
(1) the effect of the periodically oscillating force 
on the resonant frequency shift and dissipation signal in FM-AFM
and (2) the detail analysis of the electrostatic force 
by explicitly taking into account the effect of the tip oscillation.

We validated the theory by fitting with the experimental $\df$-$\Vbias$
and $g$-$\Vbias$ curves in both $1\omega$ and $2\omega$D-KPFM cases.
We experimentally showed the equivalence of $1\omega$D-KPFM and AM-KPFM
and that of $2\omega$D-KPFM and FM-KPFM in terms of their sensitivity 
to electrostatic force and electrostatic force gradient, respectively.
We demonstrated that $1\omega$D-KPFM requires a significantly smaller
ac voltage amplitude (a few tens of mV) than $2\omega$D-KPFM and FM-KPFM.
Even though the potential contrast obtained by $1\omega$D-KPFM is about 
two times smaller than that by $2\omega$D-KPFM, 
the use of the small ac voltage is of great advantage for characterizing 
materials whose properties are sensitive to the externally applied electric 
field such as semiconductors.
The operations in $1\omega$ and $2\omega$D-KPFM can be switched easily 
to take advantage of both features at the same location on a sample.

\bibliographystyle{spphys}

\begin{thebibliography}{10}
\providecommand{\url}[1]{{#1}}
\providecommand{\urlprefix}{URL }
\expandafter\ifx\csname urlstyle\endcsname\relax
  \providecommand{\doi}[1]{DOI \discretionary{}{}{}#1}\else
  \providecommand{\doi}{DOI \discretionary{}{}{}\begingroup
  \urlstyle{rm}\Url}\fi

\bibitem{Nonnenmacher91}
M.~Nonnenmacher, M.P. O’Boyle, H.K. Wickramasinghe, Appl. Phys. Lett.
  \textbf{58}(25), 2921 (1991).
\newblock \doi{10.1063/1.105227}.
\newblock \urlprefix\url{http://aip.scitation.org/doi/10.1063/1.105227}

\bibitem{Kikukawa96}
A.~Kikukawa, S.~Hosaka, R.~Imura, Rev. Sci. Instrum. \textbf{67}(4), 1463
  (1996).
\newblock \doi{10.1063/1.1146874}.
\newblock \urlprefix\url{http://aip.scitation.org/doi/10.1063/1.1146874}

\bibitem{Sommerhalter1999}
C.~Sommerhalter, T.W. Matthes, T.~Glatzel, A.~J\"{a}ger-Waldau, M.C.
  Lux-Steiner, Appl. Phys. Lett. \textbf{75}(2), 286 (1999).
\newblock \urlprefix\url{http://aip.scitation.org/doi/abs/10.1063/1.124357}

\bibitem{Zerweck05}
U.~Zerweck, C.~Loppacher, T.~Otto, S.~Grafstrom, L.M. Eng, Phys. Rev. B
  \textbf{71}(12), 125424 (2005).
\newblock \urlprefix\url{https://link.aps.org/doi/10.1103/PhysRevB.71.125424}

\bibitem{Kitamura98}
S.~Kitamura, M.~Iwatsuki, Appl. Phys. Lett. \textbf{72}(24), 3154 (1998).
\newblock \doi{10.1063/1.121577}.
\newblock \urlprefix\url{http://aip.scitation.org/doi/10.1063/1.121577}

\bibitem{Schumacher2015}
Z.~Schumacher, Y.~Miyahara, L.~Aeschimann, P.~Gr\"{u}tter, Beilstein Journal of
  Nanotechnology \textbf{6}, 1450 (2015).
\newblock \doi{10.3762/bjnano.6.150}.
\newblock
  \urlprefix\url{http://www.beilstein-journals.org/bjnano/content/6/1/150}

\bibitem{Burke09a}
S.A. Burke, J.M. LeDue, Y.~Miyahara, J.M. Topple, S.~Fostner, P.~Grutter,
  Nanotechnology \textbf{20}(26), 264012 (2009).
\newblock \doi{10.1088/0957-4484/20/26/264012}.
\newblock \urlprefix\url{http://iopscience.iop.org/0957-4484/20/26/264012}

\bibitem{Miyahara2015}
Y.~Miyahara, J.~Topple, Z.~Schumacher, P.~Grutter, Phys.~Rev.~Applied
  \textbf{4}(5), 054011 (2015).
\newblock \doi{10.1103/PhysRevApplied.4.054011}.
\newblock
  \urlprefix\url{http://link.aps.org/doi/10.1103/PhysRevApplied.4.054011}

\bibitem{Miyahara2017b}
Y.~Miyahara, P.~Grutter, Applied Physics Letters \textbf{110}(16), 163103
  (2017).
\newblock \doi{10.1063/1.4981937}.
\newblock \urlprefix\url{http://aip.scitation.org/doi/10.1063/1.4981937}

\bibitem{Fukuma2004}
T.~Fukuma, K.~Kobayashi, H.~Yamada, K.~Matsushige, Rev. Sci. Instrum.
  \textbf{75}(11), 4589 (2004).
\newblock \urlprefix\url{http://aip.scitation.org/doi/10.1063/1.1805291}

\bibitem{Nomura2007}
H.~Nomura, K.~Kawasaki, T.~Chikamoto, Y.J. Li, Y.~Naitoh, M.~Kageshima,
  Y.~Sugawara, Appl.~Phys.~Lett. \textbf{90}(3), 033118 (2007).
\newblock \doi{10.1063/1.2432281}.
\newblock \urlprefix\url{http://aip.scitation.org/doi/10.1063/1.2432281}

\bibitem{Albrecht1991}
T.R. Albrecht, P.~Grutter, D.~Horne, D.~Rugar, Journal of Applied Physics
  \textbf{69}(2), 668 (1991).
\newblock \doi{10.1063/1.347347}.
\newblock \urlprefix\url{http://aip.scitation.org/doi/10.1063/1.347347}

\bibitem{Durig1997}
U.~D\"{u}rig, H.R. Steinauer, N.~Blanc, Journal of Applied Physics
  \textbf{82}(1997), 3641 (1997).
\newblock \doi{10.1063/1.365726}.
\newblock \urlprefix\url{http://aip.scitation.org/doi/10.1063/1.365726}

\bibitem{Labuda2012a}
A.~Labuda, K.~Kobayashi, Y.~Miyahara, P.~Gr{\"{u}}tter, Rev.~Sci.~Instrum.
  \textbf{83}(May), 053703 (2012).
\newblock \doi{10.1063/1.4712286}.
\newblock \urlprefix\url{http://dx.doi.org/10.1063/1.4712286}

\bibitem{Labuda2011}
A.~Labuda, Y.~Miyahara, L.~Cockins, P.~Gr\"{u}tter, Physical Review B
  \textbf{84}(12), 125433 (2011).
\newblock \doi{10.1103/PhysRevB.84.125433}.
\newblock \urlprefix\url{http://link.aps.org/doi/10.1103/PhysRevB.84.125433}

\bibitem{Holscher2001}
H.~H\"{o}lscher, B.~Gotsmann, W.~Allers, U.~Schwarz, H.~Fuchs, R.~Wiesendanger,
  Phys. Rev. B \textbf{64}(7), 075402 (2001).
\newblock \doi{10.1103/PhysRevB.64.075402}.
\newblock \urlprefix\url{http://link.aps.org/doi/10.1103/PhysRevB.64.075402}

\bibitem{Kantorovich2004}
L.N. Kantorovich, T.~Trevethan, Phys.~Rev.~Lett. \textbf{93}(23), 236102
  (2004).
\newblock \doi{10.1103/PhysRevLett.93.236102}.
\newblock \urlprefix\url{http://link.aps.org/doi/10.1103/PhysRevLett.93.236102}

\bibitem{Sader2005}
J.E. Sader, T.~Uchihashi, M.J. Higgins, A.~Farrell, Y.~Nakayama, S.P. Jarvis,
  Nanotechnology \textbf{16}(3), S94 (2005).
\newblock
  \urlprefix\url{http://iopscience.iop.org/article/10.1088/0957-4484/16/3/018}

\bibitem{Miyahara2017}
Y.~Miyahara, A.~Roy-Gobeil, P.~Grutter, Nanotechnology \textbf{28}(6), 064001
  (2017).
\newblock \doi{10.1088/1361-6528/aa5261}.
\newblock \urlprefix\url{http://doi.org/10.1088/1361-6528/aa5261}

\bibitem{Fukuma05}
T.~Fukuma, M.~Kimura, K.~Kobayashi, K.~Matsushige, H.~Yamada, Review of
  Scientific Instruments \textbf{76}(5), 53704 (2005).
\newblock \urlprefix\url{http://aip.scitation.org/doi/10.1063/1.1896938}

\bibitem{Zahl2010}
P.~Zahl, T.~Wagner, R.~M{\"{o}}ller, A.~Klust,
  J.~Vac.~Sci.~Technol.~B~Nanotechnol.~Microelectron. \textbf{28}(3), C4E39
  (2010).
\newblock \doi{10.1116/1.3374719}.
\newblock \urlprefix\url{http://avs.scitation.org/doi/10.1116/1.3374719}

\bibitem{Diesinger2008}
H.~Diesinger, D.~Deresmes, J.P. Nys, T.~M{\'{e}}lin, Ultramicroscopy
  \textbf{108}(8), 773 (2008).
\newblock \doi{10.1016/j.ultramic.2008.01.003}.
\newblock
  \urlprefix\url{http://linkinghub.elsevier.com/retrieve/pii/S0304399108000132}

\bibitem{Melin2011}
T.~M\'{e}lin, S.~Barbet, H.~Diesinger, D.~Th\'{e}ron, D.~Deresmes, Rev. Sci.
  Instrum. \textbf{82}(3), 036101 (2011).
\newblock \doi{10.1063/1.3516046}.
\newblock \urlprefix\url{http://aip.scitation.org/doi/10.1063/1.3516046}

\bibitem{Sadewasser03}
S.~Sadewasser, M.~Lux-Steiner, Phys. Rev. Lett. \textbf{91}(26), 1 (2003).
\newblock \doi{10.1103/PhysRevLett.91.266101}.
\newblock \urlprefix\url{http://link.aps.org/doi/10.1103/PhysRevLett.91.266101}

\bibitem{Miyahara2012}
Y.~Miyahara, L.~Cockins, P.~Grutter, in \emph{Kelvin Probe Force Microscopy},
  ed. by S.~Sadewasser, T.~Glatzel (Springer Berlin Heidelberg, 2012), chap.~9,
  pp. 175--199.
\newblock \doi{10.1007/978-3-642-22566-6-9}.
\newblock
  \urlprefix\url{http://www.springer.com/materials/surfaces+interfaces/book/978-3-642-22565-9?changeHeader}

\bibitem{Denk91}
W.~Denk, D.W. Pohl, Applied Physics Letters \textbf{59}(17), 2171 (1991).
\newblock \doi{10.1063/1.106088}.
\newblock
  \urlprefix\url{http://scitation.aip.org/content/aip/journal/apl/59/17/10.1063/1.106088}

\bibitem{Melcher07}
J.~Melcher, S.~Hu, A.~Raman, Appl. Phys. Lett. \textbf{91}(5), 53101 (2007).
\newblock \urlprefix\url{http://aip.scitation.org/doi/10.1063/1.2767173}

\bibitem{Glatzel2003a}
T.~Glatzel, S.~Sadewasser, M.~Lux-Steiner, Appl. Surf. Sci. \textbf{210}(1-2),
  84 (2003).
\newblock \doi{10.1016/S0169-4332(02)01484-8}.
\newblock
  \urlprefix\url{http://linkinghub.elsevier.com/retrieve/pii/S0169433202014848}

\bibitem{Hochwitz1996}
T.~Hochwitz, A.K. Henning, C.~Levey, C.~Daghlian,
  J.~Vac.~Sci.~Technol.~B~Nanotechnol.~Microelectron. \textbf{14}(1), 457
  (1996).
\newblock \doi{10.1116/1.588494}.
\newblock
  \urlprefix\url{http://scitation.aip.org/content/avs/journal/jvstb/14/1/10.1116/1.588494}

\bibitem{Jacobs1999}
H.O. Jacobs, H.F. Knapp, A.~Stemmer, Review of Scientific Instruments
  \textbf{70}(3), 1756 (1999).
\newblock \doi{10.1063/1.1149664}.
\newblock \urlprefix\url{http://aip.scitation.org/doi/10.1063/1.1149664}

\bibitem{Strassburg2005a}
E.~Strassburg, A.~Boag, Y.~Rosenwaks, Review of Scientific Instruments
  \textbf{76}(8), 083705 (2005).
\newblock \doi{10.1063/1.1988089}.
\newblock
  \urlprefix\url{http://scitation.aip.org/content/aip/journal/rsi/76/8/10.1063/1.1988089}

\bibitem{Cohen2013}
G.~Cohen, E.~Halpern, S.U. Nanayakkara, J.M. Luther, C.~Held, R.~Bennewitz,
  A.~Boag, Y.~Rosenwaks, Nanotechnology \textbf{24}(29), 295702 (2013).
\newblock \doi{10.1088/0957-4484/24/29/295702}.
\newblock
  \urlprefix\url{http://iopscience.iop.org/article/10.1088/0957-4484/24/29/295702}

\bibitem{Cockins2010a}
L.~Cockins, Y.~Miyahara, S.D. Bennett, A.a. Clerk, S.~Studenikin, P.~Poole,
  A.~Sachrajda, P.~Grutter, Proc. Natl. Acad. Sci. U. S. A. \textbf{107}(21),
  9496 (2010).
\newblock \doi{10.1073/pnas.0912716107}.
\newblock \urlprefix\url{http://www.pnas.org/content/107/21/9496}

\end{thebibliography}

\printindex
\end{document}